\documentclass[letterpaper,12pt,notoc]{JHEP3}
\def\pd{\partial}
\def\mc{\mathcal}

\usepackage{graphicx}
\usepackage{amsmath}
\usepackage{amssymb}

\preprint{ \hbox{}\hfill arXiv:1312.4275}

\title{New $N=5,6$, 3D gauged supergravities and holography}
\author{Auttakit Chatrabhuti$^{a, \, b}$, Parinya
Karndumri$^{a,\, b}$ and Boonpithak Ngamwatthanakul$^a$\\
$^a$String Theory and Supergravity Group, Department of Physics, \\
Faculty of Science, Chulalongkorn University, 254 Phayathai Road, Pathumwan, Bangkok 10330, Thailand \\
$^b$Thailand Center of Excellence in Physics, CHE, Ministry of
Education, Bangkok 10400, Thailand
\\
E-mail:\email{auttakit@sc.chula.ac.th},
\email{parinya.ka@hotmail.com},\email{boonpithak@gmail.com} }

\abstract{We study $N=5$ gauged supergravity in three dimensions
with compact, non-compact and non-semisimple gauge groups. The
theory under consideration is of Chern-Simons type with
$USp(4,k)/USp(4)\times USp(k)$ scalar manifold. We classify possible
semisimple gauge groups of the $k=2,4$ cases and identify some of
their critical points. A number of supersymmetric $AdS_3$ critical
points are found, and holographic RG flows interpolating between
these critical points are also investigated. As one of our main
results, we consider a non-semisimple gauge group $SO(5)\ltimes
\mathbf{T}^{10}$ for the theory with $USp(4,4)/USp(4)\times USp(4)$
scalar manifold. The resulting theory describes $N=5$ gauged
supergravity in four dimensions reduced on $S^1/\mathbb{Z}_2$ and
admits a maximally supersymmetric $AdS_3$ critical point with
$Osp(5|2,\mathbb{R})\times Sp(2,\mathbb{R})$ superconformal
symmetry. We end the paper with the construction of $SO(6)\ltimes
\mathbf{T}^{15}$ gauged supergravity with $N=6$ supersymmetry. The
theory admits a half-supersymmetric domain wall as a vacuum solution
and may be obtained from an $S^1/\mathbb{Z}_2$ reduction of $N=6$
gauged supergravity in four dimensions.} 
\begin{document}
\section{Introduction}
The duality between scalars and vectors together with the
non-propagating nature of supergravity fields in three dimensions
make three dimensional gauged supergravity substantially differs
from its higher dimensional analogue. On one hand, only
matter-coupled supergravity has propagating degrees of freedom in
terms of scalars and spin-$\frac{1}{2}$ fields. Accordingly, the
matter-coupled theory takes the form of a supersymmetric non-linear
sigma model coupled to supergravity. On the other hand, recasting
vectors to scalars, making the U-duality symmetry manifest, seems to
create a trouble in any attempt to gauge the theory since the vector
fields accompanying for the gauging are missing.
\\
\indent Special to three dimensions, vector fields can enter the
gauged Lagrangian via Chern-Simons (CS) terms as opposed to the
conventional Yang-Mills (YM) kinetic terms. Since CS terms do not
lead to additional degrees of freedom, any number of gauge fields,
or equivalently the dimension of the gauge group, can be introduced
provided that the gauge group is a proper subgroup of the global
symmetry group and consistent with supersymmetry. This gives rise to
a very rich structure of gauged supergravity in three dimensions
\cite{nicolai1, nicolai2,nicolai3, N8, dewit}.
\\
\indent Additionally, the Chern-Simons form of gauged supergravity
raises another difficulty namely the embedding of the resulting
gauged theory in higher dimensions. This is due to the fact that all
theories obtained from conventional dimensional reductions are of
Yang-Mills form. It has been, however, shown that Yang-Mills gauged
supergravity is on-shell equivalent to Chern-Simons gauged theory
with a non-semisimple gauge group \cite{csym}. Up to now, there are
many attempts to embed three dimensional gauged supergravity in
higher dimensions and in string/M theory. These results would give
rise to new string theory backgrounds with fluxes as well as new
D-brane configurations \cite{gaugeSUGRA_flux}. However, it has been
pointed out recently in \cite{non_SUGRA_background} that there might
exist supersymmetric string backgrounds which are not captured by
gauged supergravities.
\\
\indent The rich structure and embedding in string/M theory aside,
gauged supergravity proves to be a very useful tool in the AdS/CFT
correspondence \cite{maldacena}. AdS$_3$/CFT$_2$ correspondence can
provide more insight not only to the AdS/CFT correspondence,
including its generalizations such as the Domain Wall/Quantum Field
Theory (DW/QFT) correspondence, but also to black hole physics
\cite{Strominger_note, krause lecture}. In holographic RG flows,
$AdS_3$ vacua and domain walls interpolating between them
interpreted as RG flows in the dual two dimensional field theories
are of particular interest, see \cite{freedman} for a thorough
review. The deformations of a strongly coupled field theory can be
understood in this framework. Some gauged supergravities do not
admit a maximally supersymmetric $AdS_3$ but a half-supersymmetric
domain wall as a vacuum solution. This class of gauged
supergravities will be useful in the context of the DW/QFT
correspondence
\cite{DW/QFT_townsend,correlator_DW/QFT,Skenderis_DW/QFT}.
\\
\indent In this work, we further explore the structure of gauged
supergravity in three dimensions with $N=5,6$ supersymmetry. We
begin with a study of compact and non-compact gaugings of the $N=5$
theory with scalar manifolds $USp(4,2)/USp(4)\times USp(2)$ and
$USp(4,4)/USp(4)\times USp(4)$. We will identify some supersymmetric
$AdS_3$ critical points and study the associated RG flow solutions.
This could be useful in AdS$_3$/CFT$_2$ correspondence although the
embedding in higher dimensions is presently not known. The result is
similar to supersymmetric RG flows studied in \cite{bs, gkn, AP, N6}
and in higher dimensions such as recent solutions of new maximal
gauged supergravity in four dimensions given in
\cite{DW_in_new4DgaugedSUGRA}.
\\
\indent We then move to non-semisimple gaugings of the $N=5$ theory
containing 16 scalars encoded in $USp(4,4)/USp(4)\times USp(4)$
coset manifold with $SO(5)\ltimes \mathbf{T}^{10}$ gauge group. The
gauge group is embedded in the global symmetry group $USp(4,4)$.
According to \cite{csym}, the resulting theory is equivalent to
$SO(5)$ YM gauged supergravity. The latter might be obtained by a
reduction of $N=5$, $SO(5)$ gauged supergravity in four dimensions
on $S^1/\mathbb{Z}_2$ as pointed out in \cite{DW_from_N10}. The
theory may also be embedded in $N=10$, $SO(5)\ltimes
\mathbf{T}^{10}$ gauged supergravity via the embedding of the global
symmetry group $USp(4,4)\subset E_{6(-14)}$. The theory admits a
maximally supersymmetric $AdS_3$ vacuum and provides another example
of three dimensional gauged supergravities with known higher
dimensional origin.
\\
\indent We finally turn to non-semisimple gauging of $N=6$ theory
with $SU(4,4)/S(U(4)\times U(4))$ scalar manifold. The global
symmetry $SU(4,4)$ contains an $SO(6)\ltimes \mathbf{T}^{15}$
subgroup that can be consistently gauged. Similar to $N=5$ theory,
this theory is equivalent to $SO(6)$ YM gauged supergravity and
could be obtained by an $S^1/\mathbb{Z}_2$ reduction of $N=6$ gauged
supergravity in four dimensions. Unlike $N=5$ theory, the theory
admits only a half-supersymmetric domain wall as a vacuum solution.
\\
\indent The paper is organized as follow. We give the construction
of $N=5$ theory in section \ref{N5_theory}. Relevant information and
related formulae for general gauged supergravity in three dimensions
are collected in appendix \ref{general_construction}. Vacua of
compact and non-compact gauge groups are given in section
\ref{compact} and \ref{non_compact}, respectively. Section
\ref{flow} deals with some examples of RG flows between critical
points previously identified. Non-semisimple gaugings of $N=5$ and
$N=6$ theories are constructed in sections \ref{N5_nonsemi} and
\ref{N6_nonsemi}, respectively. The maximally supersymmetric $AdS_3$
of $N=5$ theory and a $\frac{1}{2}$-BPS domain wall of the $N=6$
theory are explicitly given in these sections. We end the paper with
some conclusions and discussions. Appendices \ref{generators} and
\ref{potential_SO4_k2} contain the explicit form of the relevant
generators used in the main text as well as the scalar potential for
$SO(4)\times USp(2)$ gauging in $N=5$ theory.
\section{$N=5$ gauged supergravity in three
dimensions}\label{N5_theory} In $N=5$ three dimensional gauged
supergravity, scalar fields are described in term of
$USp(4,k)/USp(4)\times USp(k)$ coset manifold with dimensionality
$4k$. The R-symmetry is given by $USp(4)\sim SO(5)_R$. All
admissible gauge groups are embedded in the global symmetry group
$USp(4,k)$. In this paper, we will consider only the $k=2$ and $k=4$
cases.
\\
\indent We first introduce $USp(4,k)$ generators constructed from a
compact group $USp(4+k)$ via the Weyl unitarity trick. In order to
make contact with the $N=6$ theory with global symmetry group
$SU(4,k)$ studied in section \ref{N6_nonsemi}, we will construct the
$USp(4+k)$ generators by figuring out the $USp(4+k)$ subgroup of
$SU(4+k)$, directly. The latter is generated by the well-known
generalized Gell-Mann matrices given in, for example, \cite{stancu}.
We will denote $USp(4+k)$ generators by $J_{i}$ given explicitly in
appendix \ref{generators}. The $SO(5)_R$ R-symmetry generators,
labeled by a pair of anti-symmetric indices $T^{IJ}=-T^{JI}$, can be
identified as follow
\begin{eqnarray}
T^{12}&=&\frac{1}{\sqrt{2}}\left( J_3-J_6 \right),\qquad
T^{13}=-\frac{1}{\sqrt{2}}\left( J_1+J_4 \right),\qquad
T^{23}=\frac{1}{\sqrt{2}}\left( J_2-J_5 \right),\nonumber \\
T^{34}&=&\frac{1}{\sqrt{2}}\left( J_3+J_6 \right),\qquad
T^{14}=\frac{1}{\sqrt{2}}\left( J_2+J_5 \right),\qquad
T^{24}=\frac{1}{\sqrt{2}}\left( J_1-J_4 \right),\nonumber \\
T^{15}&=&-J_9,\qquad T^{25}=-J_{10},\qquad  T^{35}=J_8,\qquad
T^{45}=J_7\, .
\end{eqnarray}
\indent The non-compact generators $Y^A$ are identified by
\begin{eqnarray}
Y^{1}&=&iJ_{14},\qquad Y^{2}=iJ_{15},\qquad Y^{3}=iJ_{16},\qquad Y^{4}=iJ_{17}, \nonumber \\
Y^{5}&=&iJ_{18},\qquad Y^{6}=iJ_{19},\qquad Y^{7}=iJ_{20},\qquad Y^{8}=iJ_{21}, \nonumber \\
Y^{9}&=&iJ_{25},\qquad Y^{10}=iJ_{26},\qquad Y^{11}=iJ_{27},\qquad Y^{12}=iJ_{28}, \nonumber \\
Y^{13}&=&iJ_{29},\qquad Y^{14}=iJ_{30},\qquad Y^{15}=iJ_{31},\qquad
Y^{16}=iJ_{32}\, .
\end{eqnarray}
For $k=2$ case with 8 scalars, the associated non-compact generators
are given by the first 8 generators, $Y^A$ with $A=1,\ldots, 8$.
\\
\indent Admissible gauge groups are completely characterized by the
symmetric gauge invariant embedding tensor $\Theta_{\mc{MN}}$,
$\mc{M},\mc{N}=1,\ldots , \textrm{dim} \, G$. Viable gaugings are
defined by the embedding tensor satisfying two constraints. The
first constraint is quadratic in $\Theta$ and given by
\begin{equation}
\Theta_{\mathcal{PL}}f^{\mathcal{KL}}{}_{(\mathcal{M}}\Theta_{\mathcal{N)K}}=0
\end{equation}
ensuring that a given gauge group $G_0$ is a proper subgroup of $G$.
The other constraint due to supersymmetry takes the form of a
projection condition
\begin{equation}
\mathbb{P}_\boxplus T^{IJ,KL}=0
\end{equation}
where the T-tensor $T^{IJ,KL}$ is given by the moment map of the
embedding tensor
\begin{equation}
T^{IJ,KL} \equiv
\mathcal{V}^{\mathcal{M}\,IJ}\Theta_{\mathcal{M}\mathcal{N}}\mathcal{V}^{\mathcal{N}\,KL}\,
.
\end{equation}
The $\boxplus$ denotes the Riemann tensor-like representation of
$SO(N)_R$. For symmetric scalar manifolds of the form $G/H$, the
$\mc{V}$ maps can be obtained from the coset representative, see
appendix \ref{general_construction}, and the constraint can be
written in the form
\begin{equation}
\mathbb{P}_{R_0}\Theta_{\mathcal{MN}}=0\, .
\end{equation}
The representation $R_0$ of $G$ contains the $\boxplus$
representation of $SO(N)_R$.
\\
\indent We are now in a position to study gaugings of $N=5$
supergravity. We will treat compact and non-compact gauge groups
separately.
\section{Compact gauge groups}\label{compact}
In this section, we explore $N=5$ gauged supergravity with compact
gauge groups. The gauge groups are subgroup of $USp(4)\times USp(k)$
and takes the form $SO(p) \times SO(5-p) \times USp(k)$, $p=5,4,3$.
\\
\indent The $SO(p) \times SO(5-p)$ part is embedded in $SO(5)_R$ as
$\mathbf{5}\rightarrow
(\mathbf{p},\mathbf{1})+(\mathbf{1},\mathbf{5-p})$. The
corresponding embedding tensor is identified in \cite{dewit} and
takes the form
\begin{equation}
\Theta_{IJ,KL}=\theta \delta^{KL}_{IJ}+\delta_{[I[K}\Xi_{L]J]}
\end{equation}
where
\begin{equation}
\Xi_{IJ}=\left\{\begin{array}{c}
                 2\left(1-\frac{p}{5}\right)\delta_{IJ}, \qquad I\leq p \\
                 -\frac{2p}{5}\delta_{IJ},\qquad I>p
               \end{array}
\right ., \qquad \theta=\frac{2p-5}{5}\, .
\end{equation}
The full embedding tensor for $SO(p) \times SO(5-p) \times USp(k)$
is given by
\begin{equation}
\Theta=g_1\Theta_{SO(p)\times SO(5-p)}+g_2 \Theta _{USp(k)}
\end{equation}
with two independent coupling constants. $\Theta_{USp(k)}$ is given
by the Killing form of $USp(k)$. Together with the explicit form of
the coset representative, the scalar potential is completely
determined by the embedding tensor.

\subsection{The $k=2$ case}
In this case, the theory contains 8 scalars parametrized by
$USp(4,2)/USp(4)\times USp(2)$ coset space. The full 8-dimensional
manifold can be conveniently parametrized by the Euler angles of
$SO(5) \times USp(2)\sim USp(4)\times USp(2)$. The details of the
parametrization can be found in \cite{exceptional coset}, and the
application to $SU(n,m)/S(U(n)\times U(m))$ coset can be found in
\cite{N6}.
\subsubsection{$SO(5)\times USp(2)$ gauging}\label{secSO5k2}
With $USp(4)\times USp(2)$ Euler angles, the full
$USp(4,2)/USp(4)\times USp(2)$ coset can be parametrized by the
coset representative
\begin{equation}
L=e^{a_1X_1}e^{a_2X_2}e^{a_3X_3}e^{a_4J_7}e^{a_5J_8}e^{a_6J_9}e^{a_7J_{15}}e^{bY^{7}}\label{L_Euler}
\end{equation}
 where $X_i$'s are defined by
\begin{equation}
X_1=\frac{1}{\sqrt{2}}(J_1-J_{11}),\qquad
X_2=\frac{1}{\sqrt{2}}(J_2-J_{12}),\qquad
X_3=\frac{1}{\sqrt{2}}(J_3-J_{13}).\label{X_i}
\end{equation}
The resulting scalar potential is
\begin{eqnarray}\label{eq:VSO5k2}
V&=&\frac{1}{32} \left[64 \left(g_2^2-12 g_1^2+4 g_1 g_2\right) \cosh b -1076 g_1^2-180 g_1 g_2-45 g_2^2 \right.\nonumber\\
&&\left.-4 \left(52 g_1^2+20 g_1 g_2+5 g_2^2\right) \cosh(2 b)+(2
g_1+g_2)^2 \cosh(4 b)\right].\label{V_SO5_k2}
\end{eqnarray}
Note that the scalar fields associated to the gauge generators do
not appear in the potential due to gauge invariance. We find some
critical points as shown in table \ref{tab:SO5k2}. $V_0$ is the
value of the potential at each critical point. Unbroken
supersymmetry is denoted by $(n_-,n_+)$ where $n_-$ and $n_+$
correspond to the number of supersymmetry in the dual two
dimensional CFT. In three dimensional language, they correspond to
the numbers of negative and positive eigenvalues of $A_1^{IJ}$
tensor. As reviewed in appendix \ref{general_construction}, these
eigenvalues, $\pm \tilde{\alpha}$, satisfy $V_0=-4\tilde{\alpha}^2$.
Since, in our convention, the $AdS_3$ radius is given by
$L=\frac{1}{\sqrt{-V_0}}$, we also have a relation
$L=\frac{1}{2|\tilde{\alpha}|}$.
\begin{table}
\begin{center}
\begin{tabular}{|c|c|c|c|c|}
  \hline
  &$b$ &  $V_0$ & unbroken & unbroken \\
  & &   & SUSY & gauge symmetry \\\hline
  I &$0$ & $-64g_1^2$ & $(5,0)$ & $SO(5)\times USp(2)$ \\
  II &$\cosh^{-1}\left[ \frac{g_2-2g_1}{2g_1+g_2}\right]$
  &  $-\frac{64g_1^2(g_1+g_2)^2}{(2g_1+g_2)^2}$ & $(4,0)$ & $USp(2)\times USp(2)$ \\
  III &$\cosh^{-1}\left[ \frac{6g_1+g_2}{2g_1+g_2}\right]$ &
  $-\frac{64g_1^2(3g_1+g_2)^2}{(2g_1+g_2)^2}$ & $(1,0)$ & $USp(2)\times USp(2)$ \\
  \hline
\end{tabular}
  \caption{\small Critical points of $SO(5)\times USp(2)$ gauging.}
  \label{tab:SO5k2}
\end{center}
\end{table}
\\
\indent The maximally supersymmetric critical point at
$L=\textbf{I}$ preserves the full gauge symmetry. The two
non-trivial critical points preserve $USp(2)\times USp(2)$ symmetry.
We also give the $A_1$ tensors at each critical point:
\begin{eqnarray}
A_1^\text{(I)}&=& -4g_1\mathbf{I}_{5\times 5},\nonumber \\
A_1^\text{(II)}&=&\text{diag}\left( \alpha, \alpha, \alpha, \alpha, \frac{4g_1(g_1-g_2)}{2g_1+g_2} \right).\nonumber \\
A_1^\text{(III)}&=&\text{diag}\left( \beta, \beta, \beta, \beta,
\frac{-4g_1(3g_1+g_2)}{2g_1+g_2} \right).
\end{eqnarray}
where
\begin{equation}
\alpha=\frac{-4g_1(g_1+g_2)}{2g_1+g_2},\qquad
\beta=\frac{-4g_1(5g_1+g_2)}{2g_1+g_2}\, .
\end{equation}
\indent The scalar mass spectrum at the trivial critical point is
given in the table below.
\begin{center}
\begin{tabular}{|c|c|}
  \hline
  $m^2L^2$  & $SO(5)\times USp(2)$  \\ \hline
  $-\frac{3}{4}$ &  $(\mathbf{4},\mathbf{2})$ \\\hline
\end{tabular}
\end{center}
All scalars have the same mass $m^2L^2=-\frac{3}{4}$ with $L$ being
the $AdS_3$ radius at this critical point. The full symmetry of the
background corresponds to $Osp(5|2,\mathbb{R})\times
Sp(2,\mathbb{R})$ superconformal group. Notice that in finding
critical points with constant scalars we can use the gauge symmetry
and the composite $USp(4)\times USp(k)$ symmetry to fix the scalar
parametrization as, for example, in the Euler angle parametrization.
In determining scalar masses, we need to compute scalar fluctuations
to quadratic order. In this case, only the the composite
$USp(4)\times USp(k)$ symmetry can be used since the vector fields
are set to zero, see the discussion in \cite{7D_critical_point}. The
scalar masses must accordingly be computed in the so-called unitary
gauge with the coset representative
\begin{equation}
L=\prod_{i=1}^8e^{a_iY^i}\, .
\end{equation}
\indent The mass spectrum at $(4,0)$ critical point is shown below.
\begin{center}
\begin{tabular}{|c|c|}
  \hline
  $m^2L^2$  & $USp(2)\times USp(2)$  \\ \hline
  $\frac{g_2 (2 g_1+3 g_2)}{(g_1+g_2)^2}$ &  $(\mathbf{1},\mathbf{1})$ \\\hline
$0$ &  $(\mathbf{2},\mathbf{2})+(\mathbf{1},\mathbf{3})$ \\\hline
\end{tabular}
\end{center}
And, scalar masses at $(1,0)$ critical point are as follow.
\begin{center}
\begin{tabular}{|c|c|}
  \hline
  $m^2L^2$  & $USp(2)\times USp(2)$  \\ \hline
  $\frac{(4 g_1+g_2) (10 g_1+3 g_2)}{(3 g_1+g_2)^2}$ &  $(\mathbf{1},\mathbf{1})$ \\\hline
$0$ &  $(\mathbf{2},\mathbf{2})+(\mathbf{1},\mathbf{3})$ \\\hline
\end{tabular}
\end{center}
Notice that there are seven massless Goldstone bosons corresponding
to the symmetry breaking $SO(5)\times USp(2)\rightarrow USp(2)\times
USp(2)$.

\subsubsection{$SO(4)\times USp(2)$ gauging}
We still use the same parametrization as in the previous case. The
potential in this case turns out to be much more complicated
although it dose not depend on $a_1$, $a_2$ and $a_3$. We give its
explicit form in appendix \ref{potential_SO4_k2}.  The trivial
critical point has $N=(4,1)$ supersymmetry and preserves the full
$SO(4)\times USp(2)$ symmetry. The $A_1$ tensor and scalar masses at
this point are given below.
\begin{equation}
A_1^\text{(I)}=-4g_1\text{diag}\left(1 ,1,1,1,-1 \right),
\end{equation}
\begin{center}
\begin{tabular}{|c|c|}
  \hline
  $m^2L^2$  & $SO(4)\times USp(2)\sim SU(2)\times SU(2)\times USp(2)$  \\ \hline
  $-\frac{3}{4}$ &  $(\mathbf{2},\mathbf{1},\mathbf{2})+(\mathbf{1},\mathbf{2},\mathbf{2})$ \\\hline
\end{tabular}
\end{center}
The corresponding superconformal symmetry is
$Osp(4|2,\mathbb{R})\times Osp(1|2,\mathbb{R})$.
\\
\indent Other critical points with $a_4=a_5=a_6=a_7=0$ are shown in
table \ref{tab:SO4k2}.
\begin{table}
\begin{center}
\begin{tabular}{|c|c|c|c|c|}
  \hline
  &$b$ &  $V_0$ & unbroken & unbroken \\
  & &   & SUSY & gauge symmetry \\\hline
  I &$0$ & $-64g_1^2$ & $(4,1)$ & $SO(4)\times USp(2)$ \\
  II &$\cosh^{-1}\left[ \frac{g_2-2g_1}{2g_1+g_2}\right]$
  &  $-\frac{64g_1^2(g_1+g_2)^2}{(2g_1+g_2)^2}$ & $(4,1)$ & $USp(2)\times USp(2)$ \\
  III &$\cosh^{-1}\left[ \frac{6g_1+g_2}{2g_1+g_2}\right]$ &
  $-\frac{64g_1^2(3g_1+g_2)^2}{(2g_1+g_2)^2}$ & $(0,0)$ & $USp(2)\times USp(2)$ \\
  \hline
\end{tabular}
\caption{\small Critical points of $SO(4)\times USp(2)$ gauging.}
  \label{tab:SO4k2}
\end{center}
\end{table}
Critical points II and III preserve only $USp(2)_\text{diag}\times
USp(2)$ subgroup of $SO(4)\times USp(2)$. The $USp(2)_\text{diag}$
is a diagonal subgroup of one factor in $USp(2)\times USp(2)\sim
SO(4)$ and the $USp(2)$ factor in the gauge group and is generated
by $J_1+J_{11}$,$J_2+J_{12}$ and $J_3+J_{13}$. Critical point II has
$(4,1)$ supersymmetry with the $A_1$ tensor
\begin{equation}
A_1^\text{(II)}=-\frac{4g_1(g_1+g_2)}{2g_1+g_2}\text{diag}\left( 1,
1, 1, 1, -1\right).
\end{equation}
The scalar mass spectrum is given in the table below.
\begin{center}
\begin{tabular}{|c|c|}
  \hline
  $m^2L^2$  & $USp(2)\times USp(2)$  \\ \hline
  $0$ &  $(\mathbf{1},\mathbf{3})$ \\\hline
  $\frac{g_2 (2 g_1+3 g_2)}{(g_1+g_2)^2}$ &  $(\mathbf{1},\mathbf{1})$ \\\hline
  $-\frac{g_1 g_2 (g_1+2 g_2)}{(g_1+g_2)^2 (2 g_1+g_2)}$ &  $(\mathbf{2},\mathbf{2})$ \\\hline
\end{tabular}
\end{center}
Critical point III is non-supersymmetric with scalar masses given by
\begin{center}
\begin{tabular}{|c|c|}
  \hline
  $m^2L^2$  & $USp(2)\times USp(2)$  \\ \hline
  $0$ &  $(\mathbf{1},\mathbf{3})$ \\ \hline
  $\frac{(4 g_1+g_2) (10 g_1+3 g_2)}{(3 g_1+g_2)^2}$ &  $(\mathbf{1},\mathbf{1})$ \\\hline
  $-\frac{g_1 (4 g_1+g_2) (5 g_1+2 g_2)}{(2 g_1+g_2) (3 g_1+g_2)^2}$ &  $(\mathbf{2},\mathbf{2})$ \\ \hline
\end{tabular}\, .
\end{center}
We can now check its stability by comparing the above scalar masses
with the Breitenlohner-Freedman bound $m^2L^2\geq -1$. At this
critical point, the value of $b$ is real for $g_1>0$ and $g_2>-2g_1$
or $g_1<0$ and $g_2<-2g_1$. For definiteness, we will consider the
first possibility. The mass of the singlet scalar satisfies the BF
bound for $g_1>0$ and $g_2>-3g_1$ while the mass of
$(\mathbf{2},\mathbf{2})$ scalars requires $g_2>0.21432 g_1$ for
$g_1>0$ to satisfy to BF bound. Therefore, critical point III is
stable for $g_1>0$ and $g_2>0.21432 g_1$.
\\
\indent Note that both critical points II and III contain three
massless scalars which are responsible for the symmetry breaking
$SO(4)\times USp(2)\rightarrow USp(2)\times USp(2)$.

\subsubsection{$SO(3)\times SO(2)\times USp(2)$ gauging}
Computing the scalar potential on the full 8-dimensional manifold
turns out to be very complicated even with the Euler angle
parametrization \eqref{L_Euler}. In order to make things more
tractable, we employ the technique introduced in \cite{warner} and
consider a submanifold of $USp(4,2)/USp(4)\times USp(2)$ invariant
under $U(1)_\text{diag}$ symmetry generated by $T^{12}+T^{45}$.
There are four singlets under this symmetry corresponding to the
non-compact generators
\begin{eqnarray}
X_1&=&\frac{1}{\sqrt{2}}(Y^1+Y^6),\qquad X_2=\frac{1}{\sqrt{2}}(Y^2+Y^8),\nonumber \\
X_3&=&\frac{1}{\sqrt{2}}(Y^4-Y^3),\qquad
X_4=\frac{1}{\sqrt{2}}(Y^7-Y^5).
\end{eqnarray}
The coset representative can be parametrized by
\begin{equation}
L=e^{a_1X_1}e^{a_2X_2}e^{a_3X_3}e^{a_4X_4}.\label{U1D_coset}
\end{equation}
The resulting potential is given by
\begin{eqnarray}
V&=&\frac{1}{128} \left[3+\cosh a_1  \cosh a_2  \cosh a_3  \cosh a_4 \right] \left[-2 \left(512 g_1^2+19 g_2^2\right)\right.\nonumber\\
&&\left.+\left(99 g_2^2-1024 g_1^2\right) \cosh a_1  \cosh a_2  \cosh a_3  \cosh a_4 +3 g_2^2 \cosh(2 a_1)\times \right.\nonumber\\
 &&\left.(\cosh a_1  \cosh a_2  \cosh a_3  \cosh a_4 )-2-12 g_2^2 \cosh^2 a_1 \left[\cosh(2 a_2)\right.\right.\nonumber\\
 &&\left.\left.+2 \cosh^2 a_2 \left(\cosh(2 a_3)+2 \cosh^2 a_3  \cosh(2 a_4)\right)\right]+2 g_2^2 \cosh^3 a_1\times \right.\nonumber\\
 & & \cosh a_2  \cosh a_3  \left(3 \left(\cosh(2 a_2)+2 \cosh^2 a_2  \cosh(2 a_3)\right) \cosh a_4 \right.\nonumber \\
 & &\left.\left.+4 \cosh^2a_2 \cosh^2 a_3 \cosh(3 a_4)\right)\right]. \label{V_SO3_k2}
 \end{eqnarray}
We find critical points as shown in table \ref{tab:SO3k2}. We have
given only the value of $a_1$ since, at all critical points, the
four scalars are related by $a_2=a_1$ and $a_3=a_4=0$.
\begin{table}
\begin{center}
\begin{tabular}{|c|c|c|c|c|}
  \hline
  &$a_1$ &  $V_0$ & unbroken & unbroken \\
  & &   & SUSY & gauge symmetry \\\hline
  I &$0$ & $-64g_1^2$ & $(3,2)$ & $SO(3)\times SO(2)\times USp(2)$ \\
  II &$\frac{1}{2}\ln\left[\frac{g_2-8 g_1 - 4 \sqrt{g_1 (4 g_1 - g_2)}}{g_2}\right]$
  &  $-\frac{64g_1^2(g_1-g_2)^2}{g_2^2}$ & $(2,0)$ & $U(1)\times U(1)$ \\
  III &$\frac{1}{2}\ln\left[\frac{g_2+8 g_1 - 4 \sqrt{g_1 (4 g_1 + g_2)}}{g_2}\right]$ & $-\frac{64g_1^2(g_1+g_2)^2}{g_2^2}$ & $(1,2)$ & $U(1)\times U(1)$ \\
  \hline
\end{tabular}
\caption{\small Critical points of $SO(3)\times SO(2) \times USp(2)$
gauging.}
  \label{tab:SO3k2}
\end{center}
\end{table}
As usual, when all scalars vanish, we have a maximally
supersymmetric point with $N=(3,2)$ and $SO(3)\times SO(2)\times
USp(2)$ symmetry. The corresponding $A_1$ tensor is
\begin{equation}
A_1^\text{(I)}= -4g_1\text{diag}\left(1,1,1,-1,-1 \right).
\end{equation}
This background leads to the superconformal symmetry
$Osp(3|2,\mathbb{R})\times Osp(2|2,\mathbb{R})$. The scalar masses
at this point are shown below.
\begin{center}
\begin{tabular}{|c|c|}
  \hline
  $m^2L^2$  & $SO(2)\times SO(3)\times USp(2)$  \\ \hline
  $-\frac{3}{4}$ &  $(1,\mathbf{2},\mathbf{2})+(-1,\mathbf{2},\mathbf{2})$ \\\hline
\end{tabular}
\end{center}
\indent The other two critical points preserve $U(1) \times U(1)$
symmetry. The corresponding $A_1$ tensor at these points is given by
\begin{eqnarray}
A_1^\text{(II)}&=&=\text{diag}\left( \alpha, \alpha,\beta , -\beta, -\beta \right),\nonumber \\
A_1^\text{(III)}&=&\text{diag}\left( \gamma, \gamma,-\delta , \delta
, \delta \right)
\end{eqnarray}
where
\begin{eqnarray}
\alpha &=&\frac{4g_1(g_1-g_2)}{g_2},\qquad \beta=-\frac{4g_1(g_2-3g_1)}{g_2},\nonumber \\
\gamma &=&-\frac{4g_1(3g_1+g_2)}{g_2},\qquad  \delta
=\frac{4g_1(g_1+g_2)}{g_2}\, .
\end{eqnarray}
With some normalization of the $U(1)$ charges, the scalar mass
spectra can be computed as shown in the tables below. The original
four singlets under $U(1)_{\textrm{diag}}$ correspond to one
massless and three massive modes in the tables. The
$U(1)_{\textrm{diag}}$ is given by a combination of the two $U(1)$'s
in the unbroken symmetry $U(1)\times U(1)$. Therefore, the
$(0,\pm4)$ and $(\pm 4,0)$ modes, which are singlets under one of
the two $U(1)$'s, will not be invariant under
$U(1)_{\textrm{diag}}$.
\begin{itemize}
\item $(2,0)$ point:
\begin{center}
\begin{tabular}{|c|c|}
  \hline
  $m^2L^2$  & $U(1)\times U(1)$  \\ \hline
  $0$ &  $(0,4)+(0,-4)+(4,0)+(-4,0)+(0,0)$ \\\hline
  $\frac{32 g_1^2-32 g_1 g_2+6 g_2^2}{(g_1-g_2)^2}$ &  $(0,0)$ \\\hline
  $-\frac{2 g_1 (g_1-2 g_2)}{(g_1-g_2)^2}$ &  $(-2,-2)+(2,2)$ \\\hline
\end{tabular}
\end{center}
\item $(1,2)$ point:
\begin{center}
\begin{tabular}{|c|c|}
  \hline
  $m^2L^2$  & $U(1)\times U(1)$  \\ \hline
  $0$ &  $(0,4)+(0,-4)+(4,0)+(-4,0)+(0,0)$ \\\hline
  $\frac{32 g_1^2+32 g_1 g_2+6 g_2^2}{(g_1+g_2)^2}$ &  $(0,0)$ \\\hline
  $\frac{2 g_1 (3 g_1+2 g_2)}{(g_1+g_2)^2}$ &  $(-2,-2)+(2,2)$ \\\hline
\end{tabular}
\end{center}
\end{itemize}
\subsection{The $k=4$ case}
We now consider a bigger scalar manifold
$\frac{USp(4,4)}{USp(4)\times USp(4)}$. Compact gauge groups in this
case are $SO(5) \times USp(4)$, $SO(4) \times USp(4)$ and $SO(3)
\times SO(2) \times USp(4)$. Analyzing the potential on the full
16-dimensional manifold would be very complicated. We then choose a
particular submanifold invariant under a certain subgroup of the
gauge group and study the potential on this restricted scalar
manifold as in the $SO(3)\times SO(2)\times USp(2)$ gauge group of
the previous case. The procedure is parallel to that of the $k=2$
case, so we will omit some irrelevant details particularly the
explicit form of the $A_1$ tensor at each critical point.

\subsubsection{$SO(5)\times USp(4)$ gauging}
We use the parametrization of a submanifold invariant under
$USp(2)\subset USp(4)$. There are eight singlets under this $USp(2)$
symmetry corresponding to non-compact generators of $USp(4,2)\subset
USp(4,4)$. With the Euler angle parametrization, we can write the
coset representative as
\begin{equation}
L=e^{a_1\tilde{X}_1}e^{a_2\tilde{X}_2}e^{a_3\tilde{X}_3}e^{a_4K_1}e^{a_5K_2}e^{a_6K_3}e^{a_7K_4}e^{bY^{8}}
\end{equation}
where
\begin{eqnarray}
\tilde{X}_1&=&\frac{1}{\sqrt{2}}(J_4-J_{11}),\qquad \tilde{X}_2=\frac{1}{\sqrt{2}}(J_5-J_{12}),\qquad \tilde{X}_3=\frac{1}{\sqrt{2}}(J_6-J_{13}),\nonumber \\
K_1&=&J_{31},\qquad K_2=J_{32},\qquad K_3=J_{33},\qquad K_4=J_{36}\,
.
\end{eqnarray}
The scalar potential turns out to be same as in \eqref{V_SO5_k2}.
The critical points are shown in table \ref{tab:SO5k4}.
\begin{table}
\begin{center}
\begin{tabular}{|c|c|c|c|c|}
  \hline
  &$b$ &  $V_0$ & unbroken & unbroken \\
  & &   & SUSY & gauge symmetry \\\hline
  I &$0$ & $-64g_1^2$ & $(5,0)$ & $SO(5)\times USp(4)$ \\
  II &$\cosh^{-1}\left[ \frac{g_2-2g_1}{2g_1+g_2}\right]$
  &  $-\frac{64g_1^2(g_1+g_2)^2}{(2g_1+g_2)^2}$ & $(4,0)$ & $USp(2)^3$ \\
  III &$\cosh^{-1}\left[ \frac{6g_1+g_2}{2g_1+g_2}\right]$ &
  $-\frac{64g_1^2(3g_1+g_2)^2}{(2g_1+g_2)^2}$ & $(1,0)$ & $USp(2)^3$ \\
  \hline
\end{tabular}
\caption{\small Critical points of $SO(5)\times USp(4)$ gauging.}
  \label{tab:SO5k4}
\end{center}
\end{table}
The critical points have the same structure as in the $k=2$ case but
with bigger residual symmetry. The scalar mass spectra at each
critical point are given in the tables below.
\begin{itemize}
\item $(5,0)$ point:
\begin{center}
\begin{tabular}{|c|c|}
  \hline
  $m^2L^2$  & $SO(5)\times USp(4)$  \\ \hline
  $-\frac{3}{4}$ &  $(\mathbf{4},\mathbf{4})$ \\\hline
\end{tabular}
\end{center}
\item $(4,0)$ point:
\begin{center}
\begin{tabular}{|c|c|}
  \hline
  $m^2L^2$  & $USp(2)\times USp(2)\times USp(2)$  \\ \hline
  $0$ &  $(\mathbf{2},\mathbf{2},\mathbf{1})+(\mathbf{1},\mathbf{2},\mathbf{2})+(\mathbf{1},\mathbf{3},\mathbf{1})$ \\\hline
  $\frac{g_2 (2 g_1+3 g_2)}{(g_1+g_2)^2}$ &  $(\mathbf{1},\mathbf{1},\mathbf{1})$ \\\hline
  $-\frac{4 g_1^2+8 g_1 g_2+3 g_2^2}{4 (g_1+g_2)^2}$ &  $(\mathbf{2},\mathbf{1},\mathbf{2})$ \\\hline
\end{tabular}
\end{center}
\item $(1,0)$ point:
\begin{center}
\begin{tabular}{|c|c|}
  \hline
  $m^2L^2$  & $USp(2)\times USp(2)\times USp(2)$  \\ \hline
  $0$ &  $(\mathbf{2},\mathbf{2},\mathbf{1})+(\mathbf{1},\mathbf{2},\mathbf{2})+(\mathbf{1},\mathbf{3},\mathbf{1})$ \\\hline
  $\frac{40 g_1^2+22 g_1 g_2+3 g_2^2}{(3 g_1+g_2)^2}$ &  $(\mathbf{1},\mathbf{1},\mathbf{1})$ \\\hline
  $-\frac{3 \left(12 g_1^2+8 g_1 g_2+g_2^2\right)}{4 (3 g_1+g_2)^2}$ &  $(\mathbf{2},\mathbf{1},\mathbf{2})$ \\\hline
\end{tabular}
\end{center}
\end{itemize}
Notice that the number of massless Goldstone bosons agrees with the
corresponding symmetry breaking in each case.

\subsubsection{$SO(4)\times USp(4)$ gauging}
With the same coset representative, we find the same potential as
shown in \eqref{V_SO4_k2}. The critical points with different
unbroken symmetry are shown in table \ref{tab:SO4k4}.
\begin{table}
\begin{center}
\begin{tabular}{|c|c|c|c|c|}
  \hline
  &$b$ &  $V_0$ & unbroken & unbroken \\
  & &   & SUSY & gauge symmetry \\\hline
  I &$0$ & $-64g_1^2$ & $(4,1)$ & $SO(4)\times USp(4)$ \\
  II &$\cosh^{-1}\left[ \frac{g_2-2g_1}{2g_1+g_2}\right]$
  &  $-\frac{64g_1^2(g_1+g_2)^2}{(2g_1+g_2)^2}$ & $(4,1)$ & $USp(2)^3$ \\
  III &$\cosh^{-1}\left[ \frac{6g_1+g_2}{2g_1+g_2}\right]$ &
  $-\frac{64g_1^2(3g_1+g_2)^2}{(2g_1+g_2)^2}$ & $(0,0)$ & $USp(2)^3$ \\
  \hline
\end{tabular}
\caption{\small Critical points of $SO(4)\times USp(4)$ gauging.}
  \label{tab:SO4k4}
\end{center}
\end{table}
The scalar mass spectra are given below.
\begin{itemize}
\item $(4,1)$ point:
\begin{center}
\begin{tabular}{|c|c|}
  \hline
  $m^2L^2$  & $SO(4)\times USp(2)\sim SU(2)\times SU(2)\times USp(4)$  \\ \hline
  $-\frac{3}{4}$ &  $(\mathbf{2},\mathbf{1},\mathbf{4})+(\mathbf{1},\mathbf{2},\mathbf{4})$ \\\hline
\end{tabular}
\end{center}
\item $(4,1)$ point:
\begin{center}
\begin{tabular}{|c|c|}
  \hline
  $m^2L^2$  & $USp(2)\times USp(2)\times USp(2)$  \\ \hline
  $0$ &  $(\mathbf{1},\mathbf{2},\mathbf{2})+(\mathbf{1},\mathbf{3},\mathbf{1})$ \\\hline
  $\frac{g_2 (2 g_1+3 g_2)}{(g_1+g_2)^2}$ &  $(\mathbf{1},\mathbf{1},\mathbf{1})$ \\\hline
  $-\frac{g_1 g_2 (g_1+2 g_2)}{(g_1+g_2)^2 (2 g_1+g_2)}$ &  $(\mathbf{2},\mathbf{1},\mathbf{2})$ \\\hline
  $-\frac{(2 g_1+g_2) (2 g_1+3 g_2)}{4 (g_1+g_2)^2}$ &  $(\mathbf{2},\mathbf{2},\mathbf{1})$ \\\hline
\end{tabular}
\end{center}
\item Non-supersymmetric point:
\begin{center}
\begin{tabular}{|c|c|}
  \hline
  $m^2L^2$  & $USp(2)\times USp(2)\times USp(2)$  \\ \hline
  $0$ &  $(\mathbf{1},\mathbf{2},\mathbf{2})+(\mathbf{1},\mathbf{3},\mathbf{1})$ \\\hline
  $\frac{40 g_1^2+22 g_1 g_2+3 g_2^2}{(3 g_1+g_2)^2}$ &  $(\mathbf{1},\mathbf{1},\mathbf{1})$ \\\hline
  $-\frac{3 (2 g_1+g_2) (6 g_1+g_2)}{4 (3 g_1+g_2)^2}$ &  $(\mathbf{2},\mathbf{1},\mathbf{2})$ \\\hline
  $-\frac{g_1 \left(20 g_1^2+13 g_1 g_2+2 g_2^2\right)}{(2 g_1+g_2) (3 g_1+g_2)^2}$ &  $(\mathbf{2},\mathbf{2},\mathbf{1})$ \\\hline
\end{tabular}
\end{center}
This critical point is stable for $g_1>0$ and $g_2>0.21432 g_1$.
\end{itemize}

\subsubsection{$SO(3)\times SO(2)\times USp(4)$ gauging}
In this case, we use the parametrization of $L$ as in
\eqref{U1D_coset}. The four scalars correspond to four singlets of
$USp(2)\times U(1)_{\textrm{diag}}$. The potential is the same as
\eqref{V_SO3_k2} with the critical points shown in table
\ref{tab:SO3k4}.
\begin{table}
\begin{center}
\begin{tabular}{|c|c|c|c|c|}
  \hline
  &$a_1$ &  $V_0$ & unbroken & unbroken \\
  & &   & SUSY & gauge symmetry \\\hline
  I &$0$ & $-64g_1^2$ & $(3,2)$ & $SO(3)\times SO(2)\times USp(4)$ \\
  II &$\frac{1}{2}\ln\left[\frac{g_2-8 g_1 - 4 \sqrt{g_1 (4 g_1 - g_2)} }{g_2}\right]$
  &  $-\frac{64g_1^2(g_1-g_2)^2}{g_2^2}$ & $(2,0)$ & $U(1)\times U(1)\times USp(2)$ \\
  III &$\frac{1}{2}\ln\left[\frac{g_2+8 g_1 - 4 \sqrt{g_1 (4 g_1 + g_2)}}{g_2}\right]$
  & $-\frac{64g_1^2(g_1+g_2)^2}{g_2^2}$ & $(1,2)$ & $U(1)\times U(1)\times USp(2)$ \\
  \hline
\end{tabular}
\caption{\small Critical points of $SO(3)\times SO(2) \times USp(4)$
gauging.}
  \label{tab:SO3k4}
\end{center}
\end{table}
The scalar mass spectra are given in the following tables.
\begin{itemize}
\item $(3,2)$ point:
\begin{center}
\begin{tabular}{|c|c|}
  \hline
  $m^2L^2$  & $SO(3)\times USp(4)$  \\ \hline
  $-\frac{3}{4}$ &  $(\mathbf{2},\mathbf{4})+(\mathbf{2},\mathbf{4})$ \\\hline
\end{tabular}
\end{center}
\item $(2,0)$ point:
\begin{center}
\begin{tabular}{|c|c|}
  \hline
  $m^2L^2$  & $ U(1) \times U(1)\times USp(2)$  \\ \hline
  $0$ &  $(4,0,\mathbf{1})+(-4,0,\mathbf{1})+(0,4,\mathbf{1})+(0,-4,\mathbf{1})+(0,0,\mathbf{1})$ \\
  & $+(1,-1,\mathbf{2})+(-1,1,\mathbf{2})$ \\\hline
  $\frac{32 g_1^2-32 g_1 g_2+6 g_2^2}{(g_1-g_2)^2}$ &  $(0,0,\mathbf{1})$ \\\hline
  $-\frac{2 g_1 (g_1-2 g_2)}{(g_1-g_2)^2}$ &  $(-2,-2,\mathbf{1})+(2,2,\mathbf{1})$ \\\hline
  $-\frac{4 g_1^2-8 g_1 g_2+3 g_2^2}{4 (g_1-g_2)^2}$ &  $(-1,-1,\mathbf{2})+(1,1,\mathbf{2})$ \\\hline
\end{tabular}
\end{center}
\item $(1,2)$ point:
\begin{center}
\begin{tabular}{|c|c|}
  \hline
  $m^2L^2$  & $U(1) \times U(1)\times USp(2)$  \\ \hline
  $0$ &  $(4,0,\mathbf{1})+(-4,0,\mathbf{1})+(0,4,\mathbf{1})+(0,-4,\mathbf{1})+(0,0,\mathbf{1})$ \\
  & $+(1,-1,\mathbf{2})+(-1,1,\mathbf{2})$ \\\hline
  $\frac{32 g_1^2+32 g_1 g_2+6 g_2^2}{(g_1+g_2)^2}$ &  $(0,0,\mathbf{1})$ \\\hline
  $-\frac{2 g_1 (3g_1+2 g_2)}{(g_1+g_2)^2}$ &  $(-2,-2,\mathbf{1})+(2,2,\mathbf{1})$ \\\hline
  $-\frac{4 g_1^2+8 g_1 g_2+3 g_2^2}{4 (g_1+g_2)^2}$ &  $(-1,-1,\mathbf{2})+(1,1,\mathbf{2})$ \\\hline
\end{tabular}
\end{center}
\end{itemize}
That critical points in the $k=4$ case are similar to those in the
$k=2$ case should be related to the fact that the theory with
$USp(4,2)/USp(4)\times USp(2)$ scalar manifold can be embedded in
the theory with $USp(4,4)/USp(4)\times USp(4)$ scalar manifold. We
have studied the potential on scalars which are singlets under
$USp(2)$. These singlets are precisely parametrized by non-compact
directions of $USp(4,2)\subset USp(4,4)$, the global symmetry group
of $k=2$ case. This might explain the fact that this particular
parametrization gives rise to the same potential as in the $k=2$
case. Turning on more scalars would give more interesting
structures.
\section{Non-compact gauge groups}\label{non_compact}
In this section, we classify admissible non-compact gauge groups. We
will consider the $k=2$ and $k=4$ cases separately as in the
previous section.
\subsection{The $k=2$ case}
In this case, there is only one non-compact subgroup of $USp(4,2)$
namely $USp(2,2)$. The $USp(4,2)$ itself can be gauged with the
embedding tensor given by its Killing form, but the corresponding
potential will become a cosmological constant. The subgroup of
$USp(4,2)$ that can be gauged is $USp(2)\times USp(2,2)\subset
USp(4,2)$. The embedding tensor reads
\begin{equation}
 \Theta=g_1\Theta_{USp(2)}+g_2 \Theta_{USp(2,2)}
\end{equation}
where $g_1$ and $g_2$ are two independent coupling constants.
$\Theta_{USp(2,2)}$ and $\Theta_{USp(2)}$ are given by the Killing
forms of $USp(2,2)$ and $USp(2)$, respectively.
\\
\indent Generally, scalar fields corresponding to non-compact
directions in the gauge group will drop out from the potential.
Therefore, we do not need to include them in the coset
representative. The remaining four scalars correspond to non-compact
directions of another $USp(2,2)$ in $USp(4,2)$ and can be
parametrized by the coset representative of $USp(2,2)/USp(2)\times
USp(2)$. We can use Euler angles of $USp(2)\times USp(2)$ to
parametrize the coset representative as
\begin{equation}
L=e^{a_1X_1}e^{a_2X_2}e^{a_3X_3}e^{bY^{7}}
\end{equation}
where $X_i$ are given in \eqref{X_i}. We find the following
potential
\begin{eqnarray}
V&=&\frac{1}{16} \left[8(g_1-g_2+(g_1+g_2) \cosh(b))^2 \sinh^2 b\right.\nonumber\\
&&\left.-\left(3 g_1+11 g_2+4 (g_1-g_2) \cosh b +(g_1+g_2) \cosh(2
b)\right)^2\right].
\end{eqnarray}
Some of the critical points are shown in table \ref{tab:NCk2}.
\begin{table}
\begin{center}
\begin{tabular}{|c|c|c|c|c|}
  \hline
  &$b$ &  $V_0$ & unbroken & unbroken\\
  & &   & SUSY & gauge symmetry \\\hline
  I &$0$ & $-4(g_1+g_2)^2$ & $(4,1)$ & $USp(2)^3$ \\
  II &$\cosh^{-1}\left( \frac{g_2-g_1}{g_1+g_2}\right)$
  &  $-\frac{4g_1^2(2g_1+g_2)^2}{(g_1+g_2)^2}$ & $(4,0)$ & $USp(2)\times USp(2)$ \\
  III &$\cosh^{-1}\left( -\frac{g_1+3g_2}{g_1+g_2}\right)$ &
  $-\frac{4g_1^2(2g_1+3g_2)^2}{(g_1+g_2)^2}$ & $(1,0)$ & $USp(2)\times USp(2)$ \\
  IV &$\ln(2+\sqrt{3})$ & $-\frac{1}{4}(27g_1^2+54g_1g_2+19g_2^2)$ & $(0,0)$ & $USp(2)\times USp(2)$ \\
  \hline
\end{tabular}
\caption{\small Critical points of $USp(2)\times USp(2,2)$ gauging.}
  \label{tab:NCk2}
\end{center}
\end{table}
The $A_1$ tensor at each supersymmetric critical point is given by
\begin{eqnarray}
A_1^\text{(I)}&=&(g_1+g_2)\text{diag}\left(-1,-1,-1,-1,1 \right),\nonumber \\
A_1^\text{(II)}&=&\text{diag}\left( \beta, \beta, \beta, \beta,  \frac{g_2(-2g_1+g_2)}{g_1+g_2} \right),\nonumber \\
A_1^\text{(III)}&=&\text{diag}\left( \gamma, \gamma, \gamma,
\gamma,-\frac{g_2(2g_1+3g_2)}{g_1+g_2} \right)
\end{eqnarray}
where
\begin{equation}
\beta=-\frac{g_2(2g_1+g_2)}{g_1+g_2},\qquad
\gamma=-\frac{g_2(2g_1+5g_2)}{g_1+g_2}\, .
\end{equation}
Critical point I preserves $N=(4,1)$ supersymmetry. The gauge group
is broken down to its maximal compact subgroup $USp(2)^3$. In this
symmetry breaking, the four massless Goldstone bosons correspond to
scalars associated to non-compact generators of the gauge group. The
full symmetry at this point gives the superconformal symmetry
$Osp(4|2,\mathbb{R})\times Osp(1|2,\mathbb{R})$ since the
supercharges transform under $USp(2)\times USp(2)\subset SO(5)_R$ as
$(\mathbf{2},\mathbf{2})+(\mathbf{1},\mathbf{1})$.
\\
\indent Scalar mass spectra at all critical points are given below.
\begin{itemize}
\item $(4,1)$ point:
\begin{center}
\begin{tabular}{|c|c|}
  \hline
  $m^2L^2$  & $USp(2)\times USp(2)\times USp(2)$  \\ \hline
  $0$ &  $(\mathbf{1},\mathbf{2},\mathbf{2})$ \\\hline
  $-\frac{g_1 (g_1+2 g_2)}{(g_1+g_2)^2}$ &  $(\mathbf{2},\mathbf{1},\mathbf{2})$ \\\hline
\end{tabular}
\end{center}
\item $(4,0)$ point:
\begin{center}
\begin{tabular}{|c|c|}
  \hline
  $m^2L^2$  & $USp(2)\times USp(2)$  \\ \hline
  $0$ &  $(\mathbf{2},\mathbf{2})+(\mathbf{1},\mathbf{3})$ \\\hline
  $\frac{4 g_1 (3 g_1+g_2) }{(2 g_1+g_2)^2}$ &  $(\mathbf{1},\mathbf{1})$ \\\hline
\end{tabular}
\end{center}
\item $(1,0)$ point:
\begin{center}
\begin{tabular}{|c|c|}
  \hline
  $m^2L^2$  & $USp(2)\times USp(2)$  \\ \hline
  $0$ &  $(\mathbf{2},\mathbf{2})+(\mathbf{1},\mathbf{3})$ \\\hline
  $\frac{4 (g_1+2 g_2) (3 g_1+5 g_2) }{(2 g_1+3 g_2)^2}$ &  $(\mathbf{1},\mathbf{1})$ \\\hline
\end{tabular}
\end{center}
\item Non-supersymmetric point:
\begin{center}
\begin{tabular}{|c|c|}
  \hline
  $m^2L^2$  & $USp(2)\times USp(2)$  \\ \hline
  $0$ &  $(\mathbf{2},\mathbf{2})+(\mathbf{1},\mathbf{3})$ \\\hline
  $\frac{12 (3 g_1+g_2) (3 g_1+5 g_2) }{27 g_1^2+54 g_1 g_2+19 g_2^2}$ &  $(\mathbf{1},\mathbf{1})$ \\\hline
\end{tabular}
\end{center}
\end{itemize}
At non-trivial critical points, there are additional three massless
scalars which are responsible for $USp(2)\times USp(2)\rightarrow
USp(2)_{\textrm{diag}}$ symmetry breaking. The non-supersymmetric
critical point is stable for $g_2>\frac{3}{79}(2\sqrt{210}-45)g_1$.

\subsection{The $k=4$ case}
There are three possible non-compact subgroups of $USp(4,4)$;
$USp(2,2)\times USp(2,2)$, $USp(2)\times USp(4,2)$ and $USp(2)\times
USp(2) \times USp(2,2)$. Only $USp(2,2)\times USp(2,2)$ can be
gauged with the following embedding tensor
\begin{equation}
\Theta=g_1\Theta_{USp(2,2)}+g_2 \Theta_{USp(2,2)}\, .
\end{equation}
There are two independent coupling constants $g_1$ and $g_2$, and
$\Theta_{USp(2,2)}$ is given by the Killing form of $USp(2,2)$. The
relevant 8 scalars can be parametrized by
$\left(\frac{USp(2,2)}{USp(2)\times USp(2)}\right)^{2}$ coset space
with the two $USp(2,2)$ factors different from those appearing in
the gauge group. With the Euler angle parametrization, the coset
representative reads
\begin{equation}
L=e^{a_1X_1}e^{a_2X_2}e^{a_3X_3}e^{b_1Y^{7}}e^{a_4X_4}e^{a_5X_5}e^{a_6X_6}e^{b_2Y^{16}}
\end{equation}
where
\begin{eqnarray}
X_1&=&\frac{1}{\sqrt{2}}(J_1-J_{11}),\qquad X_2=\frac{1}{\sqrt{2}}(J_2-J_{12}),\qquad  X_3=\frac{1}{\sqrt{2}}(J_3-J_{13}), \nonumber \\
X_4&=&\frac{1}{\sqrt{2}}(J_4-J_{22}),\qquad
X_5=\frac{1}{\sqrt{2}}(J_5-J_{23}),\qquad
X_6=\frac{1}{\sqrt{2}}(J_6-J_{24}).
\end{eqnarray}
The scalar potential is given by
\begin{eqnarray}
V&=&\frac{1}{16} \left[(g_1+g_2) (6+\cosh(2 b_1))-\left(4 (g_1-g_2) \cosh b_1+4 (g_2-g_1) \cosh b_2 \right.\right.\nonumber\\
&&\left.+(g_1+g_2) \cosh(2 b_2)\right)^2+8 (g_1-g_2+(g_1+g_2) \cosh(b_1))^2 \sinh^2 b_1 \nonumber\\
&&\left.+8 (g_2-g_1+(g_1+g_2) \cosh b_2 )^2 \sinh^2b_2\right].
\end{eqnarray}
We find some critical points for $b_2=0$ as shown in table
\ref{USp2_2USp2_2_critical_point}.
\begin{table}
\begin{center}
\begin{tabular}{|c|c|c|c|c|}
  \hline
  &$b_1$ &  $V_0$ & unbroken& unbroken\\
  & &   & SUSY & gauge symmetry \\\hline
  I &$0$ & $-4(g_1+g_2)^2$ & $(4,1)$ & $USp(2)^4$ \\
  II &$\cosh^{-1}\left( \frac{-g_1+g_2}{g_1+g_2}\right)$
  &  $-\frac{4g_1^2(2g_1+g_2)^2}{(g_1+g_2)^2}$ & $(4,0)$ & $USp(2)^3$ \\
  III &$\cosh^{-1}\left( \frac{-g_1-3g_2}{g_1+g_2}\right)$ &
  $-\frac{4g_1^2(2g_1+3g_2)^2}{(g_1+g_2)^2}$ & $(1,0)$ & $USp(2)^3$ \\
  IV &$\cosh^{-1} 2$ & $-\frac{1}{4}(27g_1^2+54g_1g_2+19g_2^2)$ & $(0,0)$ & $USp(2)^3$ \\
  \hline
\end{tabular}
\caption{\small Critical points of $USp(2,2)\times USp(2,2)$
gauging.} \label{USp2_2USp2_2_critical_point}
\end{center}
\end{table}
Scalar masses at all critical points are given below.
\begin{itemize}
\item $(4,1)$ point:
\begin{center}
\begin{tabular}{|c|c|}
  \hline
  $m^2L^2$  & $USp(2)\times USp(2)\times USp(2)\times USp(2)$  \\ \hline
  $0$ &  $(\mathbf{1},\mathbf{2},\mathbf{2},\mathbf{1})+(\mathbf{2},\mathbf{1},\mathbf{1},\mathbf{2})$ \\\hline
  $-\frac{g_2 (2 g_1+g_2)}{(g_1+g_2)^2}$ &  $(\mathbf{1},\mathbf{2},\mathbf{1},\mathbf{2})$ \\\hline
  $-\frac{g_1 (g_1+2 g_2)}{(g_1+g_2)^2}$ &  $(\mathbf{2},\mathbf{1},\mathbf{2},\mathbf{1})$ \\\hline
\end{tabular}
\end{center}
\item $(4,0)$ point:
\begin{center}
\begin{tabular}{|c|c|}
  \hline
  $m^2L^2$  & $USp(2)\times USp(2)\times USp(2)$  \\ \hline
  $0$ &  $(\mathbf{2},\mathbf{2},\mathbf{1})+(\mathbf{2},\mathbf{1},\mathbf{2})+(\mathbf{3},\mathbf{1},\mathbf{1})$ \\\hline
  $\frac{4 g_1 (3 g_1+g_2)}{(2 g_1+g_2)^2}$ &  $(\mathbf{1},\mathbf{1},\mathbf{1})$ \\\hline
  $-\frac{(g_1+g_2) (3 g_1+g_2)}{(2 g_1+g_2)^2}$ &  $(\mathbf{1},\mathbf{2},\mathbf{2})$ \\\hline
\end{tabular}
\end{center}
\item $(1,0)$ point:
\begin{center}
\begin{tabular}{|c|c|}
  \hline
  $m^2L^2$  & $USp(2)\times USp(2)\times USp(2)$  \\ \hline
  $0$ &  $(\mathbf{2},\mathbf{2},\mathbf{1})+(\mathbf{2},\mathbf{1},\mathbf{2})+(\mathbf{3},\mathbf{1},\mathbf{1})$ \\\hline
  $\frac{4 \left(3 g_1^2+11 g_1 g_2+10 g_2^2\right)}{(2 g_1+3 g_2)^2}$ &  $(\mathbf{1},\mathbf{1},\mathbf{1})$ \\\hline
  $-\frac{3 \left(g_1^2+4 g_1 g_2+3 g_2^2\right)}{(2 g_1+3 g_2)^2}$ &  $(\mathbf{1},\mathbf{2},\mathbf{2})$ \\\hline
\end{tabular}
\end{center}
\item Non-supersymmetry point:
\begin{center}
\begin{tabular}{|c|c|}
  \hline
  $m^2L^2$  & $USp(2)\times USp(2)\times USp(2)$  \\ \hline
  $0$ &  $(\mathbf{2},\mathbf{2},\mathbf{1})+(\mathbf{2},\mathbf{1},\mathbf{2})+(\mathbf{3},\mathbf{1},\mathbf{1})$ \\\hline
  $\frac{12 (3 g_1+g_2) (3 g_1+5 g_2)}{27 g_1^2+54 g_1 g_2+19 g_2^2}$ &  $(\mathbf{1},\mathbf{1},\mathbf{1})$ \\\hline
  $-\frac{24 g_2 (3 g_1+g_2)}{27 g_1^2+54 g_1 g_2+19 g_2^2}$ &  $(\mathbf{1},\mathbf{2},\mathbf{2})$ \\\hline
\end{tabular}
\end{center}
\end{itemize}
At the trivial critical point, the $SO(5)_R$ R-symmetry is broken to
$SU(2)\times SU(2)\sim USp(2)\times USp(2)$. The $N=5$ supercharges
transform under this subgroup as
$(\mathbf{2},\mathbf{2})+(\mathbf{1},\mathbf{1})$. This gives rise
to $Osp(4|2,\mathbb{R})\times Osp(1|2,\mathbb{R})$ superconformal
symmetry. As in the previous case, the non-supersymmetric point is
stable for $g_2>\frac{3}{79}(2\sqrt{210}-45)g_1$.

\section{RG flow solutions}\label{flow}
Given some $AdS_3$ critical points form the previous sections, we
now consider domain wall solutions interpolating between these
critical points. The solutions can be interpreted as RG flows
describing a perturbed UV CFT flowing to another CFT in the IR.
Since the structure of critical points in both $k=2$ and $k=4$ cases
is similar, we will consider only the flows in $k=2$ case to
simplify the algebra. The study of holographic RG flows is very
similar to those in other gauged supergravities in three dimensions
\cite{bs, gkn, AP, N6}. In this paper, we will give only examples of
RG flows in compact $SO(5)\times USp(2)$ and non-compact
$USp(2,2)\times USp(2)$ gauge groups.
\\
\indent We are interested only in supersymmetric flows connecting
two supersymmetric critical points. The solution can be found by
solving BPS equations arising from supersymmetry transformations of
fermions $\delta\psi^I_\mu$ and $\delta\chi^{iI}$ which, for
convenience, we will repeat them here from \cite{dewit}
\begin{eqnarray}
\delta\psi^I_\mu
&=&\mathcal{D}_\mu\epsilon^I+gA_1^{IJ}\gamma_\mu\epsilon^J,\nonumber\\
\delta\chi^{iI}&=&
\frac{1}{2}(\delta^{IJ}\mathbf{1}-f^{IJ})^i_{\phantom{a}j}{\mathcal{D}{\!\!\!\!/}}\phi^j\epsilon^J
-gNA_2^{JIi}\epsilon^J\label{susyvar}
\end{eqnarray}
where $\mc{D}_\mu
\epsilon^I=\left(\pd_\mu+\frac{1}{2}\omega_\mu^a\gamma_a\right)\epsilon^I$
for vanishing vector fields.
\\
\indent We now employ the standard domain wall ansatz for the metric
\begin{equation}
ds^2=e^{2A(r)}dx^2_{1,1}+dr^2\, .\label{DW_ansatz}
\end{equation}
In order to preserve Poincare symmetry in two dimensions, all fields
involving in the flow can only depend on the radial coordinate $r$
identified with an energy scale in the dual field theory. BPS
equations give rise to first order flow equations describing the
dependence of active scalars on $r$. It can be verified that setting
some of the scalars to zero satisfies their flow equations. We can
then neglect all scalars that vanish at both UV and IR points.

\subsection{An RG flow between $(5,0)$ and $(4,0)$ CFT's in $SO(5)\times USp(2)$ gauging}
The flow involves only one active scalar parametrized by the coset
representative
\begin{equation}
L=e^{b(r)Y^7}\, .\label{flow_coset1}
\end{equation}
The BPS equation from $\delta \chi^{iI}=0$ gives rise to the flow
equation
\begin{equation}
\frac{db}{dr}=\left[2 g_1-g_2+(2 g_1+g_2) \cosh b \right] \sinh b
\end{equation}
where we have used the projection condition
$\gamma_r\epsilon^I=\epsilon^I$. It is clearly seen from the above
equation that there are two critical points at $b=0$ and
$b=\cosh^{-1}\frac{g_2-2g_1}{2g_1+g_2}$. This equation can be solved
for $r$ as a function of $b$, and the solution is given by
\begin{eqnarray}
r&=&\frac{1}{8g_1g_2}\left[ 4 g_1 \ln \cosh\frac{b}{2}  - (2 g_1 +
g_2) \ln[
   2 g_1 - g_2 + (2 g_1 + g_2) \cosh b ] \right.\nonumber\\
   &&\left.+ 2 g_2 \ln \sinh \frac{b}{2} \right].\label{b_sol1}
\end{eqnarray}
The integration constant has been neglected since we can shift the
coordinate $r$ to remove it.
\\
\indent The variation $\delta \psi^{I}_{\mu}=0$ gives another
equation for $A(r)$
\begin{eqnarray}
\frac{dA }{dr}&=&\frac{1}{4} \left[4 g_2 \cosh b-22 g_1-3 g_2-8 g_1 \cosh b \right.\nonumber\\
&&\left.-2 g_1 \cosh(2 b )-g_2 \cosh(2 b )\right]
\end{eqnarray}
or, in term of $b$,
\begin{equation}
\frac{dA}{db}=-\frac{\left[22 g_1+3 g_2+(8 g_1-4 g_2) \cosh b +(2
g_1+g_2) \cosh(2 b)\right] \text{csch} b }{8 g_1-4 g_2+4 (2 g_1+g_2)
\cosh b }\, .
\end{equation}
This equation is readily solved and gives $A$ as a function of $b$
\begin{eqnarray}
A&=&\frac{1}{g_2}\left[(g_1+g_2) \ln\left[2 g_1-g_2+(2 g_1+g_2) \cosh b \right]-(2 g_1+g_2) \ln \cosh \frac{b}{2} \right.\nonumber\\
&&\left.-2 g_2 \ln \sinh \frac{b}{2} \right].
\end{eqnarray}
The additive integration constant can be absorbed by scaling
$x^{0,1}$ coordinates. It can be verified that equation
$\delta\psi^I_r=0$ gives the Killing spinors of the unbroken
supersymmetry $\epsilon^I=e^{\frac{A}{2}}\epsilon_0^I$ as usual,
with constant spinors $\epsilon_0^I$ satisfying
$\gamma_r\epsilon_0^I=\epsilon_0^I$.
\\
\indent Linearizing equation \eqref{b_sol1} near the UV point
$b\approx 0$, we find
\begin{equation}
b(r)\sim e^{4g_1r}= e^{-\frac{r}{2L_{UV}}},\qquad
L_{UV}=\frac{1}{8|g_1|}.
\end{equation}
We have set $g_1<0$ to identify $r \rightarrow \infty $ as the UV
point. The above behavior indicates that from a general result, see
for example \cite{freedman}, the flow is driven by a relevant
operator of dimension $\Delta =\frac{3}{2}$.
\\
\indent Near the IR point, we find
\begin{equation}
b(r)\sim e^{-\frac{8g_1 g_2 r}{2g_1+g_2}}= e^{\frac{g_2
r}{(g_1+g_2)L_{IR}}},\qquad
L_{IR}=-\frac{2g_1+g_2}{8g_1(g_1+g_2)}>0\, .
\end{equation}
The reality condition for $b_{IR}$ requires $g_2>-2g_1$ for $g_1<0$.
From the above equation, we find $\frac{g_2}{g_2+g_1}>0$, so in the
IR the operator becomes irrelevant with dimension
$\Delta_{IR}=\frac{3g_2+2 g_2}{g_1+g_2}$. This value of
$\Delta_{IR}$ precisely gives the correct mass square
$m^2L^2_{IR}=\frac{g_2(2g_1+3g_2)}{(g_1+g_2)^2}$ given before.
\\
\indent The ratio of the central charges is computed to be
\begin{equation}
\frac{c_{UV}}{c_{IR}}=\frac{L_{UV}}{L_{IR}}=\sqrt{\frac{V_{0IR}}{V_{0UV}}}=\frac{g_1+g_2}{2
g_1 +g_2}>1
\end{equation}
satisfying the holographic c-theorem for $g_1<0$ and $g_2>-2 g_1$.

\subsection{An RG flow between $(5,0)$ and $(1,0)$ CFT's in $SO(5)\times USp(2)$ gauging}
We then study another RG flow interpolating between $(5,0)$ and
$(1,0)$ critical points. The coset representative is sill given by
\eqref{flow_coset1}. Similar to the previous case, we obtain the
following flow equations
\begin{eqnarray}
\frac{db}{dr}&=&\left[6 g_1+g_2-(2 g_1+g_2)  \cosh b \right] \sinh b ,\nonumber \\
\frac{dA }{dr}&=&\frac{1}{4}\left[3 g_2-10 g_1-4 (6 g_1+g_2) \cosh b
+(2 g_1+g_2) \cosh(2 b )\right].\label{eq_x}
\end{eqnarray}
The first equation gives a solution
\begin{eqnarray}
r&=&-\frac{1}{8 g_1 (4 g_1 + g_2)}\left[4 g_1 \ln \cosh\frac{b}{2}+(2 g_1+g_2) \ln\left[(2 g_1+g_2)\cosh b \right.\right.\nonumber\\
&&\left.\left.-6 g_1-g_2\right]-2 (4 g_1+g_2) \ln \sinh
\frac{b}{2}\right].
\end{eqnarray}
\indent We can rewrite the second equation of \eqref{eq_x} as
\begin{equation}
\frac{dA}{db}=\frac{\left[10 g_1 - 3 g_2 +
   4 (6 g_1 + g_2) \cosh b  - (2 g_1 + g_2) \cosh(2 b)\right] \text{csch} b }{ 4 (2 g_1 + g_2) \cosh b -4 (6 g_1 + g_2)}
\end{equation}
whose solution can be found to be
\begin{eqnarray}
A&=&\frac{1}{4 g_1 + g_2}\left[ (3 g_1 + g_2) \ln\left((2 g_1 + g_2)
\cosh b -6 g_1 -
    g_2\right) \right.\nonumber\\
   &&\left. -(2 g_1 + g_2) \ln
   \cosh \frac{b}{2} - 2 (4 g_1 + g_2) \ln \sinh \frac{b}{2} \right].
\end{eqnarray}
\indent The fluctuation around $b=0$ behaves as
\begin{equation}
b(r)\sim e^{4g_1r}= e^{-\frac{r}{2L_{UV}}},
L_{UV}=\frac{1}{8|g_1|}\, .
\end{equation}
As in the previous case, we have chosen $g_1<0$ to make the UV point
corresponds to $r\rightarrow \infty$. From the above equation, the
flow is again driven by a relevant operator of dimension
$\Delta_{UV}=\frac{3}{2}$. Near the IR point, $b(r)$ becomes
\begin{equation}
b(r)\sim e^{-\frac{8g_1 (4 g_1+g_2) r}{2g_1+g_2}}= e^{\frac{(4
g_1+g_2) r}{(3 g_1+g_2)L_{IR}}},\qquad
L_{IR}=-\frac{2g_1+g_2}{8g_1(3 g_1+g_2)}.
\end{equation}
We can verify that $b_{IR}$ is real for $g_1<0$ and $g_2<-2 g_1$,
the operator becomes irrelevant in the IR with dimension
$\Delta_{IR}=\frac{10 g_1+3 g_2}{3 g_1+g_2}$. The ratio of the
central charges is given by
\begin{equation}
\frac{c_{UV}}{c_{IR}}=\frac{3 g_1+g_2}{2 g_1 +g_2}>1,\qquad
\textrm{for}\qquad g_1<0\,\, \textrm{and}\,\, g_2<-2g_1\, .
\end{equation}

\subsection{An RG flow between $(4,1)$ and $(4,0)$ CFT's in $USp(2)\times USp(2,2)$ gauging}
We next consider RG flows between critical points of non-compact
$USp(2)\times USp(2,2)$ gauge group. We will not give a
non-supersymmetric flow to critical point IV in table \ref{tab:NCk2}
in this paper. It can be studied in the same procedure as
\cite{deger} and \cite{Kandumri_nonSUSYflow}. Like in the compact
case, it is consistent to truncate the full scalar manifold to a
single scalar parametrized by
\begin{equation}
L=e^{b(r)Y^7}\, .
\end{equation}
\indent The variation $\delta \chi^{iI}=0$ gives
\begin{equation}
\frac{db }{dr}=(g_1 - g_2 + (g_1 + g_2)\cosh b ) \sinh b
\end{equation}
which is solved by the solution
\begin{eqnarray}
r&=&\frac{1}{4g_1g_2}\left[2 g_2 \ln \sinh \frac{b}{2}  +2 g_1 \ln \cosh \frac{b}{2}  \right.\nonumber\\
   &&\left. \phantom{\frac{1}{2}}- (g_1 + g_2) \ln\left[
   g_1 - g_2 + (g_1 + g_2) \cosh b \right] \right].
\end{eqnarray}
The equation from $\delta \psi^{I}_{\mu}=0$ reads
\begin{equation}
\frac{dA}{dr}=-2\left[ g_2+g_1 \cosh^{4} \frac{b }{2} +g_2 \sinh^{4}
\frac{b }{2} \right].
\end{equation}
The solution for $A$ as a function of $b$ can be found as in the
previous cases. The result is given by
\begin{eqnarray}
A&=&\frac{1}{2 g_1}\left[ (2 g_1 + g_2) \ln\left[
   g_1 - g_2 + (g_1 + g_2) \cosh b\right] -4 g_1 \ln \cosh \frac{b}{2}  \right.\nonumber\\
   &&\left.- 2 (g_1 + g_2) \ln \sinh \frac{b}{2}
\right].
\end{eqnarray}
Near the UV point, the $b$ solution becomes
\begin{equation}
b(r)\sim e^{2 g_1r}= e^{\frac{g_1r}{(g_1+g_2)L_{UV}}},\qquad
L_{UV}=\frac{1}{2(g_1+g_2)}\, .
\end{equation}
$b_{IR}$ is real for $g_1<0$ and $g_2>-g_1$. With this range,
$-\frac{g_1}{g_1+g_2}<1$. The flow is then driven by a relevant
operator of dimension $\Delta=\frac{3g_1+2g_2}{g_1+g_2}<2$. At the
IR point, we find the asymptotic behavior
\begin{equation}
b(r)\sim e^{-\frac{4g_1 g_2 r}{g_1+g_2}}= e^{\frac{2 g_2
r}{|2g_1+g2|L_{IR}}},\qquad  L_{IR}=\frac{g_1+g_2}{2|g_1(2g_1+g_2)|}
\end{equation}
corresponding to an irrelevant operator of dimension
$\Delta=\frac{2g_2}{|2g_1+g2|}+2$.
\\
\indent Finally, the ratio of the central charges is given by
\begin{equation}
\frac{c_{UV}}{c_{IR}}=\frac{|g_1(2g_1+g_2)|}{(g_1+g_2)^2}\, .
\end{equation}

\subsection{An RG flow between $(4,1)$ and $(1,0)$ CFT's in $USp(2)\times USp(2,2)$ gauging}
As a final flow solution, we quickly investigate a solution
interpolating between $(4,1)$ and $(1,0)$ critical points. The flow
equations are given by
\begin{eqnarray}
\frac{db}{dr}&=&-\left[g_1 + 3 g_2 + (g_1 + g_2) \cosh b \right] \sinh b, \\
\frac{dA }{dr}&=&\frac{1}{4}\left[3 g_1 - 5 g_2 + 4 (g_1 + 3 g_2)
\cosh b + (g_1 + g_2) \cosh(2 b)\right].
\end{eqnarray}
The corresponding solutions take the form
\begin{eqnarray}
r&=&-\frac{1}{4g_2(g_1+2g_2)}\left[(g_1 + g_2) \ln\left[
   g_1 + 3 g_2 + (g_1 + g_2) \cosh b \right] \right.\nonumber\\
&&\left. + 2 g_2 \ln \sinh \frac{b}{2} -2 (g_1 + 2 g_2) \ln \cosh \frac{b}{2}\right],\\
A&=&\frac{1}{2 (g_1+2g_2)}\left[(2 g_1 + 3 g_2) \ln\left[
   g_1 + 3 g_2+ (g_1 + g_2) \cosh b \right] \phantom{\frac{1}{2}}\right.\nonumber\\
   &&\left.-4 (g_1 + 2 g_2) \ln \cosh \frac{b}{2} - 2 (g_1 + g_2) \ln \sinh \frac{b}{2}
\right].
\end{eqnarray}
The fluctuations near the UV and IR points are given by
\begin{eqnarray}
& &b(r)\sim e^{-2( g_1+2g_2)r}= e^{\frac{(g_1+2 g_2)r}{(g_1+g_2)L_{UV}}},\qquad L_{UV}=-\frac{1}{2(g_1+g_2)},\\
& &b(r)\sim e^{-\frac{4g_2(g_1+2g_2) r}{g_1+g_2}}=
e^{\frac{2g_2(g_1+2 g_2 )r}{|g_1(2g_1+3g2)|L_{IR}}},\qquad
L_{IR}=-\frac{(g_1+g_2)}{2|g_1(2g_1+3g_2)|}\, .
\end{eqnarray}
We have chosen a particular range of $g_1$ and $g_2$ namely $g_1<0$
and $-\frac{g_1}{2}<g_2<-g_1$ for which $g_1+g_2<0$. The flow is
driven by a relevant operator of dimension
$\Delta=\frac{3g_1+4g_2}{g_1+g_2}$. In the IR, the operator becomes
irrelevant with dimension $\Delta=\frac{2g_2}{|2g_1+g2|}+2$.
\\
\indent The ratio of the central charges for this flow is
\begin{equation}
\frac{c_{UV}}{c_{IR}}=\frac{|g_1(2g_1+3g_2)|}{(g_1+g_2)^2}\, .
\end{equation}

\section{$N=5$, $SO(5)\ltimes \mathbf{T}^{10}$ gauged supergravity}\label{N5_nonsemi}
\begin{center}
\begin{figure}[!h] \centering
\includegraphics[width=0.5\textwidth, bb = 0 0 250 200 ]{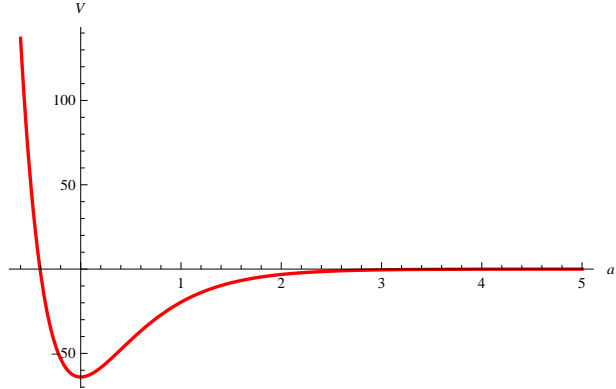}\\
\caption{The scalar potential of $N=5$, $SO(5)\ltimes
\mathbf{T}^{10}$ gauged supergravity for $SO(5)$ singlet scalar with
$g=1$.} \label{fig1}
\end{figure}
\end{center}

In this section, we consider non-semisimple gauge groups in the form
of $G_0\ltimes \mathbf{T}^{\textrm{dim}\, G_0}$ in which $G_0$ is a
semisimple group. $\mathbf{T}^{\textrm{dim}\, G_0}$ constitutes a
translational symmetry with $\textrm{dim}\, G_0$ commuting
generators transforming in the adjoint representation of $G_0$. We
consider the $k=4$ case with $USp(4,4)$ global symmetry that admits
a non-semisimple subgroup $SO(5)\ltimes \mathbf{T}^{10}$.
\\
\indent A general embedding of $G_0\ltimes
\mathbf{T}^{\textrm{dim}\, G_0}$ group is described by the embedding
tensor of the form \cite{csym}
\begin{equation}
\Theta=g_1 \Theta_{\textrm{ab}}+g_2 \Theta_{\textrm{bb}}\,
.\label{Theta_nonsemi}
\end{equation}
We have used the notation of \cite{csym} in denoting the semisimple
and translational parts by $\textrm{a}$ and $\textrm{b}$,
respectively. The absence of $\textrm{aa}$ coupling plays a key role
in the equivalence of this theory and the Yang-Mills gauged
supergravity with $G_0$ gauge group.
\\
\indent The next task is to identify $SO(5)\ltimes \mathbf{T}^{10}$
generators. The semisimple $SO(5)$ is identified with the diagonal
subgroup of $SO(5)\times SO(5) \sim USp(4)\times USp(4)\subset
USp(4,4)$. The corresponding generators are given by
\begin{equation}\label{eq:Td}
J^{ij}=T^{ij}+\tilde{T}^{ij}, \qquad i,j=1,2,...,5\, .
\end{equation}
$T^{ij}$ are the $SO(5)$ R-symmetry generators, and $\tilde{T}^{ij}$
are generators of $USp(4)$. The translational generators are
constructed from a combination of $T^{ij}-\tilde{T}^{ij}$ and
non-compact generators. The 16 scalars transform as
$(\mathbf{4},\mathbf{4})$ under $SO(5)\times SO(5)$. They
accordingly transform as $\mathbf{1}+\mathbf{5}+ \mathbf{10}$ under
$SO(5)_\textrm{diag}$. Scalars in the $\mathbf{10}$ representation
will be part of the $\mathbf{T}^{10}$ generators which are given by
\begin{equation}\label{eq:Tl}
t^{ij}=T^{ij}-\tilde{T}^{ij}+  \tilde{Y}^{ij}, \qquad
i,j=1,2,...,5\, .
\end{equation}
The explicit form of $\tilde{T}^{ij}$ and $\tilde{Y}^{ij}$ is given
in appendix \ref{generators}.
\\
\indent In the present case, supersymmetry allows for any value of
$g_1$ and $g_2$. Therefore, the embedding tensor contains two
independent coupling constants. We begin with the scalar potential
computed on the $SO(5)_\textrm{diag}$ singlet scalar. The above
decomposition gives one singlet under this $SO(5)$. We end up with a
simple coset representative
\begin{equation}
L=e^{a(Y^{7}+Y^{16})}\, .
\end{equation}
This results in the potential
\begin{equation}
V=-64g_1e^{-3a}\left(3e^{a}g_1+2g_2\right)\, .
\end{equation}
The existence of a maximally supersymmetric critical point at
$L=\mathbf{I}$ requires $g_2=-g_1$. This is the same as in $N=4,8$
gauged supergravities \cite{henning_N8AdS3_S3,N4_gauging}. With this
condition and $g_1$ denoted by $g$, the potential becomes
\begin{equation}
V=-64g^2e^{-3a}\left(3e^a-2\right).
\end{equation}
Clearly, the only one critical point is given by $a=0$ with
$V_0=-64g^2$ and $N=(5,0)$ supersymmetry. This critical point is a
minimum of the potential as can be seen from Figure \ref{fig1}. The
vacuum is very similar to the $AdS_3$ vacuum found in $N=16$,
$SO(4)\times SO(4)\ltimes (\mathbf{T}^{12},\hat{\mathbf{T}}^{34})$
gauged supergravity studied in \cite{Hohm_henning}. The singlet has
a positive mass square $m^2L^2=3$ as expected for a minimum point.
In the dual CFT with superconformal symmetry
$Osp(5|2,\mathbb{R})\times Sp(2,\mathbb{R})$, this scalar
corresponds to an irrelevant operator of dimension $\Delta = 3$. The
full scalar masses are given below.
\begin{center}
\begin{tabular}{|c|c|}
  \hline
  $m^2L^2$ & $SO(5)$ \\ \hline
  $3$ & $\mathbf{1}$ \\
  $3$ & $\mathbf{5}$ \\
  $0$ & $\mathbf{10}$ \\
  \hline
\end{tabular}
\end{center}
The ten massless scalars accompany for the symmetry breaking
$SO(5)\ltimes \mathbf{T}^{10}\rightarrow SO(5)$ at the vacuum.
\\
\indent To find other critical points, we reduce the residual
symmetry of the scalar submanifold to $SO(3)\subset SO(5)$ under
which the 16 scalars transform as $(\mathbf{2}+
\mathbf{2})\times(\mathbf{2}+ \mathbf{2})=4\times
(\mathbf{1}+\mathbf{3})$. There are four singlets which can be
parametrized by the coset representative
\begin{equation}
L=e^{a_1Y^4}e^{a_2Y^7}e^{a_3Y^9}e^{a_4Y^{16}}\, .
\end{equation}
The resulting potential turns out to be very complicated. We,
therefore, will not attempt to do the analysis of this potential in
the present work.
\section{$N=6$, $SO(6) \ltimes \mathbf{T}^{15}$ gauged supergravity}\label{N6_nonsemi}
In this section, we consider non-semisimple gauge groups of $N=6$
theory. Compact and non-compact gauge groups in this theory together
with their vacua and holographic RG flows have been studied in
\cite{N6}.
\\
\indent We are interested in $N=6$ gauged supergravity with
$\frac{SU(4,4)}{S(U(4)\times U(4))}$ scalar manifold. Most of our
conventions here are parallel to those used in \cite{N6}. The global
symmetry $SU(4,4)$ contains a non-semisimple subgroup $SO(6)\ltimes
\mathbf{T}^{15}$. Similar to $N=5$ theory, the $SO(6)$ part is given
by the diagonal subgroup of $SO(6)\times SO(6)\sim SU(4)\times SU(4)
\subset SU(4,4)$. The 32 scalars transform as
$(\mathbf{4},\mathbf{\bar{4}}) +(\mathbf{\bar{4}},\mathbf{4})$ under
$SU(4)\times SU(4)$. Under $SO(6)_{\rm diag}$, they transform as
\begin{eqnarray}
(\mathbf{4}\times\mathbf{\bar{4}})+(\mathbf{\bar{4}} \times
\mathbf{4})= \mathbf{1}+\mathbf{15}+\mathbf{1}+\mathbf{15}.
\end{eqnarray}
The adjoint representations $\mathbf{15}$'s will be used to
construct the translational generators $\mathbf{T}^{15}$. The full
$SO(6)\ltimes \mathbf{T}^{15}$ generators are given in appendix
\ref{generators}.
\\
\indent The embedding tensor is still given by
\eqref{Theta_nonsemi}, but in this case, the linear constraint
$\mathbb{P}_{R_0}\Theta=0$ requires $g_2=0$ similar to $N=16,10,8$
theories \cite{nicolai3,DW_from_N10,DW3D}. The above decomposition
gives two singlet scalars under $SO(6)$ part of the gauge group.
They correspond to non-compact generators
\begin{eqnarray}
Y_{s1}&=&\frac{1}{2}(Y^1+Y^{11}+Y^{21}+Y^{31}),\\ \nonumber
Y_{s2}&=&\frac{1}{2}(Y^2+Y^{12}+Y^{22}+Y^{32}).
\end{eqnarray}
Accordingly, the coset representative can be parametrized by
\begin{equation}\label{eq:cosetN6}
L=e^{\sqrt{2}b_1Y_{s1}}e^{\sqrt{2}b_2Y_{s2}}
\end{equation}
where we have chosen a particular normalization for later
convenience. The potential is, with $g=g_1$, given by
\begin{equation}
V=-224g^2 \left(\cosh b_1 \cosh b_2- \sinh b_2\right)^2\, .
\end{equation}
The above potential does not admit any critical points, so the
vacuum should be a half-supersymmetric domain wall. In the rest of
this section, we will find this domain wall solution.
\\
\indent The supersymmetry transformations $\delta\psi^I_\mu$ and
$\delta\chi^{iI}$ together with the domain wall ansatz
\eqref{DW_ansatz} give rise to the following BPS equations
\begin{eqnarray}
b_{1}'&=&8g\textrm{sech}b_2\sinh b_1,\label{b1_eq}\\
b_{2}'&=&-8g\left(\cosh{b_2}-\cosh{b_1}\sinh{b_2}\right),\label{b2_eq}\\
A'&=&-16g\left(\cosh{b_1}\cosh{b_2}-\sinh{b_2}\right)\label{A_eq}
\end{eqnarray}
where $'$ denotes $\frac{d}{dr}$. Equation \eqref{b1_eq} is readily
solved by setting $b_1=0$. Equation \eqref{b2_eq} now becomes
\begin{equation}
b_{2}'=-8ge^{-b_2}\, .
\end{equation}
The solution is given by
\begin{equation}\label{b2_sol}
b_{2}=\ln \left(-8gr+c_1\right)
\end{equation}
where $c_1$ is an integration constant. With $b_{1}=0$ and $b_2$
given by \eqref{b2_sol}, equation \eqref{A_eq} becomes
\begin{equation}
A'=\frac{-16g}{c_1-8gr}
\end{equation}
whose solution is easily found to be
\begin{equation}
A=2\ln\left(-8gr+c_{1}\right)+c_{2}
\end{equation}
with another integration constant $c_{2}$. The two integration
constants are not relevant because we can shift the coordinate $r$
rescale $x^{0,1}$ to remove them. As in other domain wall solutions,
the metric can be written in the form of a warped $AdS_3$ as
\begin{equation}
ds^2=\frac{1}{(8 g )^4\rho^2}\left(\frac{ dx_{1,1}^2
+d\rho^2}{\rho^2}\right)
\end{equation}
where $\rho=-\frac{1}{(8 g)^2 r}$.
\section{Conclusions and discussions}\label{conclusion}
In this paper, we have classified compact and non-compact gauge
groups of $N=5$ gauged supergravity in three dimensions with
$USp(4,2)/USp(4)\times USp(2)$ and $USp(4,4)/USp(4)\times USp(4)$
scalar manifolds. We have also identified a number of supersymmetric
$AdS_3$ vacua in each gauging and studied some examples of
supersymmetric RG flows interpolating between these vacua in both
compact and non-compact gauge groups. All of the solutions can be
analytically found, and the flows describe deformations by relevant
operators. They would be useful to the study of AdS$_3$/CFT$_2$
correspondence such as the computation of correlation functions in
the dual field theory similar to that studied in
\cite{Henning_2Dcorrelator}.
\\
\indent Among our main results, we have constructed $N=5$,
$SO(5)\ltimes \mathbf{T}^{10}$ gauged supergravity. The theory is
equivalent to $N=5$ Yang-Mills gauged supergravity and could be
obtained from $S^1/\mathbb{Z}_2$ reduction of $N=5$ gauged
supergravity in four dimensions as pointed out in
\cite{DW_from_N10}. The theory admits a maximally supersymmetric
$AdS_3$ vacuum which should be dual to a superconformal field theory
with $Osp(5|2,\mathbb{R})\times Sp(2,\mathbb{R})$ superconformal
symmetry. We have also given all of the scalar masses at this
vacuum. It is interesting to further study the scalar potential of
this theory in order to find other critical points as well as the
associated RG flow solutions. This could give some insight to the
deformations in the dual CFT.
\\
\indent Similar construction has then been extended to $N=6$ gauged
supergravity with $SU(4,4)/S(U(4)\times U(4))$ scalar manifold. The
resulting theory is $N=6$ gauged supergravity with $SO(6)\ltimes
\mathbf{T}^{15}$ gauge group. Like $N=5$ theory, this is equivalent
to $SO(6)$ Yang-Mills gauged supergravity and should be obtained
from $S^1/\mathbb{Z}_2$ reduction of $N=6$ gauged supergravity in
four dimensions. This has also been pointed out in
\cite{DW_from_N10} in which the spectrum of the $S^1$ reduction of
four dimensional $N=6$ gauged supergravity has been given. The
theory admits a half-supersymmetric domain wall vacuum rather than a
maximally supersymmetric $AdS_3$. We have also given the domain wall
solution. This solution provides another example of domain walls in
three dimensional gauged supergravity similar to the solutions of
\cite{DW_from_N10, DW3D} and might be useful in the study of DW/QFT
correspondence.
\\
\indent The above non-semisimple gaugings are of importance for
embedding the theories in higher dimensions. With the full embedding
at hand, any solutions in a three dimensional framework, which are
usually easier to find than higher dimensional ones, can be uplifted
to string/M theory in which a full geometrical interpretation can be
made. Other attempts to embed Chern-Simons gauged supergravities in
three dimensions can be found in \cite{henning_N8AdS3_S3,
N4_gauging, Hohm_henning, KK, henning_ADSCMT, EOC_PKD3}. In many
cases, the precise reduction ansatz from ten or eleven dimensions
remains to be done.
\acknowledgments This work is partially supported by Thailand Center
of Excellence in Physics through the ThEP/CU/2-RE3/12 project. P.
Karndumri is also supported by Chulalongkorn University through
Ratchadapisek Sompote Endowment Fund under grant GDNS57-003-23-002
and The Thailand Research Fund (TRF) under grant TRG5680010.

\appendix
\section{Useful formulae}\label{general_construction}
For conveniences, we collect useful formulae used throughout this
paper. The detailed discussion can be found in \cite{dewit}. All of
our discussions involve symmetric scalar manifolds of the form
$G/H$. The $G$ generators are denoted by
$t^{\mc{M}}=(T^{IJ},T^\alpha,Y^A)$ in which $T^{IJ}$ and $T^\alpha$
are $SO(N)\times H'$ generators and $Y^A$ are non-compact
generators. In the present cases, we have $H'=USp(k)$ for $N=5$ and
$H'=U(k)$ for $N=6$ theories, respectively. $SO(N)$ is the
R-symmetry.
\\
\indent The coset manifold, consisting of $d$ scalars $\phi^i$,
$i=1,\ldots, d=\textrm{dim}\, (G/H)$, can be described by a coset
representative $L$ transforming by left- and right-multiplications
of $G$ and $H$. Some useful relations are given by
\begin{eqnarray}
L^{-1}t^\mathcal{M}L&=&\frac{1}{2}\mathcal{V}^{\mathcal{M}IJ}T^{IJ}+\mathcal{V}^\mathcal{M}_{\phantom{as}\alpha}T^\alpha+
\mathcal{V}^\mathcal{M}_{\phantom{sa}A}Y^A,\label{cosetFormula}\\
L^{-1} \partial_i L&=& \frac{1}{2}Q^{IJ}_i T^{IJ}+Q^\alpha_i
T^{\alpha}+e^A_i Y^A\, .\label{cosetFormula1}
\end{eqnarray}
The first relation gives scalar matrices $\mc{V}$ used in defining a
moment map while the second gives $SO(N)\times H'$ composite
connections, $Q^{IJ}$ and $Q^\alpha$, and the vielbein on the
manifold $G/H$, $e^A_i$. Accordingly, the metric on the scalar
manifold is defined by
\begin{equation}
g_{ij}=e^A_ie^B_j\delta_{AB},\qquad i, j, A, B =1,\ldots, d\, .
\end{equation}
\indent The embedding tensor determines the fermionic mass-like
terms and the scalar potential via the T-tensor defined by
\begin{equation}
T_{\mathcal{A}\mathcal{B}}=\mathcal{V}^{\mathcal{M}}_{\phantom{as}\mathcal{A}}\Theta_{\mathcal{M}\mathcal{N}}
\mathcal{V}^{\mathcal{N}}_{\phantom{as}\mathcal{B}}\, .
\end{equation}
In the above equation, $\mc{A}$ and $\mc{B}$ label $SO(N)\times H'$
representations.
\\
\indent The $A_1^{IJ}$ and $A_{2i}^{IJ}$ tensors appearing in the
fermionic supersymmetry transformations and the scalar potential are
given in terms of linear combinations of various components of
$T_{\mc{AB}}$ by the following relations
\begin{eqnarray}
A_1^{IJ}&=&-\frac{4}{N-2}T^{IM,JM}+\frac{2}{(N-1)(N-2)}\delta^{IJ}T^{MN,MN},\nonumber\\
A_{2j}^{IJ}&=&\frac{2}{N}T^{IJ}_{\phantom{as}j}+\frac{4}{N(N-2)}f^{M(I
m}_{\phantom{as}j}T^{J)M}_{\phantom{as}m}+\frac{2}{N(N-1)(N-2)}\delta^{IJ}f^{KL\phantom{a}m}_{\phantom{as}j}T^{KL}_{\phantom{as}m}.\nonumber\\
\end{eqnarray}
The $f^{IJ}_{ij}$ tensor can be constructed from $SO(N)$ gamma
matrices or from the $SO(N)$ generators in a spinor representation.
In the present case, it is given in a flat basis by
\begin{equation}
f^{IJ}_{AB}=-2\textrm{Tr}(Y^B\left[T^{IJ},Y^A\right]).
\end{equation}
\indent The scalar potential can be computed from
\begin{equation}
V=-\frac{4}{N}\left(A_1^{IJ}A_1^{IJ}-\frac{1}{2}Ng^{ij}A_{2i}^{IJ}A_{2j}^{IJ}\right).
\end{equation}
We end this section by noting the condition for unbroken
supersymmetry. The associated Killing spinors correspond to the
eigenvectors of $A_1^{IJ}$ with eigenvalues $\pm
\sqrt{-\frac{V_0}{4}}$.

\section{Relevant generators}\label{generators}
In this appendix, we give generators of various groups used
throughout the paper.
\subsection{$N=5$ theory}
$J_i$'s are $USp(8)$ generators written in terms of generalized
Gell-Mann matrices $\lambda_i$ generating the $SU(8)$ group. They
are explicitly given by
\begin{eqnarray}
J_1&=&\frac{i \lambda_1}{\sqrt{2}},\qquad J_2=\frac{i
\lambda_2}{\sqrt{2}},\qquad
J_3=\frac{i \lambda_3}{\sqrt{2}},\nonumber \\
J_4&=&\frac{i \lambda_{13}}{\sqrt{2}},\qquad J_5=\frac{i
\lambda_{14}}{\sqrt{2}},\qquad
J_6=-\frac{i \lambda_{8}}{\sqrt{6}}+\frac{i \lambda_{15}}{\sqrt{3}},\nonumber \\
J_7&=&\frac{i \lambda_{6}}{2}+\frac{i \lambda_{9}}{2},\qquad
J_8=-\frac{i \lambda_{7}}{2}+\frac{i \lambda_{10}}{2},\qquad
J_9=\frac{i \lambda_{4}}{2}-\frac{i \lambda_{11}}{2},\nonumber \\
J_{10}&=&-\frac{i \lambda_{5}}{2}-\frac{i \lambda_{12}}{2},\qquad
J_{11}=\frac{i \lambda_{33}}{\sqrt{2}},\qquad
J_{12}=\frac{i \lambda_{34}}{\sqrt{2}},\nonumber \\
J_{13}&=&-\frac{i \lambda_{24}}{\sqrt{5}}+\sqrt{\frac{3}{10}}i
\lambda_{35},\qquad J_{14}=\frac{i \lambda_{18}}{2}+\frac{i
\lambda_{25}}{2},\qquad
J_{15}=-\frac{i \lambda_{19}}{2}+\frac{i \lambda_{26}}{2},\nonumber \\
J_{16}&=&\frac{i \lambda_{16}}{2}-\frac{i \lambda_{27}}{2},\qquad
J_{17}=\frac{i \lambda_{22}}{2}+\frac{i \lambda_{29}}{2},\qquad
J_{18}=-\frac{i \lambda_{23}}{2}+\frac{i \lambda_{30}}{2},\nonumber \\
J_{19}&=&\frac{i \lambda_{20}}{2}-\frac{i \lambda_{31}}{2},\qquad
J_{20}=-\frac{i \lambda_{17}}{2}-\frac{i \lambda_{28}}{2},\qquad
J_{21}=-\frac{i \lambda_{21}}{2}-\frac{i \lambda_{32}}{2},\nonumber \\
J_{22}&=&\frac{i \lambda_{61}}{\sqrt{2}},\qquad J_{23}=\frac{i
\lambda_{62}}{\sqrt{2}},\qquad
J_{24}=-\sqrt{\frac{3}{14}}i \lambda_{48}+\sqrt{\frac{2}{7}}i \lambda_{63},\nonumber \\
J_{25}&=&\frac{i \lambda_{38}}{2}+\frac{i \lambda_{49}}{2},\qquad
J_{26}=-\frac{i \lambda_{39}}{2}+\frac{i \lambda_{50}}{2},\qquad
J_{27}=\frac{i \lambda_{36}}{2}-\frac{i \lambda_{51}}{2},\nonumber \\
J_{28}&=&\frac{i \lambda_{42}}{2}+\frac{i \lambda_{53}}{2},\qquad
J_{29}=-\frac{i \lambda_{43}}{2}+\frac{i \lambda_{54}}{2},\qquad
J_{30}=\frac{i \lambda_{40}}{2}-\frac{i \lambda_{55}}{2},\nonumber \\
J_{31}&=&\frac{i \lambda_{46}}{2}+\frac{i \lambda_{57}}{2},\qquad
J_{32}=-\frac{i \lambda_{47}}{2}+\frac{i \lambda_{58}}{2},\qquad
J_{33}=\frac{i \lambda_{44}}{2}-\frac{i \lambda_{59}}{2},\nonumber \\
J_{34}&=&-\frac{i \lambda_{37}}{2}-\frac{i \lambda_{52}}{2},\qquad
J_{35}=-\frac{i \lambda_{41}}{2}-\frac{i \lambda_{56}}{2},\qquad
J_{36}=-\frac{i \lambda_{45}}{2}-\frac{i \lambda_{60}}{2}.
\end{eqnarray}
The $USp(6)$ generators needed for constructing $USp(4,2)$ are given
by the first 21 generators.
\\
\indent The $SO(5) \ltimes T^{10}$ generators are constructed as
follow. The $SO(5)_{\textrm{diag}}$ is generated by
$T^{ij}+\tilde{T}^{ij}$ in which
\begin{eqnarray}
\tilde{T}^{12}&=&\frac{1}{\sqrt{2}}\left( J_{13}-J_{24}
\right),\qquad \tilde{T}^{13}=-\frac{1}{\sqrt{2}}\left(
J_{11}+J_{22} \right),\qquad
\tilde{T}^{23}=\frac{1}{\sqrt{2}}\left( J_{12}-J_{23} \right),\nonumber \\
\tilde{T}^{34}&=&\frac{1}{\sqrt{2}}\left( J_{13}+J_{24}
\right),\qquad \tilde{T}^{14}=\frac{1}{\sqrt{2}}\left( J_{12}+J_{23}
\right),\qquad
\tilde{T}^{24}=\frac{1}{\sqrt{2}}\left( J_{11}-J_{22} \right),\nonumber \\
\tilde{T}^{45}&=&J_{31},\qquad \qquad \tilde{T}^{15}=-J_{33},\qquad
\qquad \tilde{T}^{25}=-J_{36},\qquad \qquad \tilde{T}^{35}=J_{32}\,
.
\end{eqnarray}
Generators $\tilde{Y}^{ij}$ in $\mathbf{T}^{10}$ are given by
\begin{eqnarray}
\tilde{Y}^{12}&=&i(J_{16}-J_{30}),\qquad
\tilde{Y}^{13}=-i(J_{14}+J_{28}),\qquad
\tilde{Y}^{23}=i(J_{15}+J_{29}),\nonumber \\
\tilde{Y}^{34}&=&i(J_{16}+J_{30}),\qquad
\tilde{Y}^{14}=i(J_{15}+J_{29}),\qquad
\tilde{Y}^{24}=i(J_{14}-J_{28}),\nonumber \\
\tilde{Y}^{45}&=&i(J_{17}+J_{25}),\qquad
\tilde{Y}^{15}=-i(J_{19}+J_{27}),\qquad
\tilde{Y}^{25}=i(J_{21}-J_{34}),\nonumber \\
\tilde{Y}^{35}&=&i(J_{18}+J_{26}).
\end{eqnarray}

\subsection{$N=6$ theory}
For conveniences, we repeat non-compact generators of $SU(4,4)$ in
terms of generalized Gell-Mann matrices, $\lambda_i$, $i=1,\ldots,
63$, given in \cite{N6}
\begin{equation}
\bar{Y}^A=\left \{ \begin{array}{c}
               \frac{1}{\sqrt{2}}c_{A+15},\qquad A=1,\ldots, 8 \\
               \frac{1}{\sqrt{2}}c_{A+16},\qquad A=9,\ldots, 16 \\
               \frac{1}{\sqrt{2}}c_{A+19},\qquad A=17,\ldots, 24 \\
               \frac{1}{\sqrt{2}}c_{A+24},\qquad A=25,\ldots, 32
             \end{array}
\right .\, .
\end{equation}
The $SO(6)_R$ R-symmetry generators are identified to be
\begin{eqnarray}
\bar{T}^{12}&=&\frac{1}{2}c_3+\frac{1}{2\sqrt{3}}c_8-\frac{1}{\sqrt{6}}c_{15},\qquad
\bar{T}^{13}=-\frac{1}{2}(c_2+c_{14}),\qquad
\bar{T}^{23}=\frac{1}{2}(c_1-c_{13}),\nonumber \\
\bar{T}^{34}&=&\frac{1}{2}c_3-\frac{1}{2\sqrt{3}}c_8+\frac{1}{\sqrt{6}}c_{15},\qquad
\bar{T}^{14}=\frac{1}{2}(c_1+c_{13}),\qquad \bar{T}^{35}=-\frac{1}{2}(c_6+c_9),\nonumber \\
\bar{T}^{56}&=&\frac{1}{\sqrt{3}}c_8+\frac{1}{\sqrt{6}}c_{15},\qquad
\bar{T}^{36}=-\frac{1}{2}(c_7+c_{10}),\qquad
\bar{T}^{24}=\frac{1}{2}(c_2-c_{14}),\nonumber \\
\bar{T}^{45}&=&\frac{1}{2}(c_7-c_{10}),\qquad
\bar{T}^{46}=\frac{1}{2}(c_9-c_6),\qquad
\bar{T}^{15}=\frac{1}{2}(c_4-c_{11}),\nonumber \\
\bar{T}^{16}&=&\frac{1}{2}(c_5-c_{12}),\qquad
\bar{T}^{25}=\frac{1}{2}(c_5+c_{12}),\qquad
\bar{T}^{26}=-\frac{1}{2}(c_4+c_{11})
\end{eqnarray}
where $c_i=-i\lambda_i$.
\\
\indent The $SO(6) \ltimes T^{15}$ generators are given by
\begin{eqnarray}
SO(6)&:&\qquad J_{\textrm{a}}^{ij}=\bar{T}^{ij}+\tilde{\bar{T}}^{ij} ,\qquad i,j =1,\ldots , 6\nonumber \\
\mathbf{T}^{15}&:&\qquad
J_{\textrm{b}}^{ij}=\bar{T}^{ij}-\tilde{\bar{T}}^{ij}+\tilde{\bar{Y}}^{ij}
\end{eqnarray}
where
\begin{eqnarray}
\tilde{\bar{T}}^{12}&=&i\left(\frac{1}{\sqrt{10}}\lambda_{24}-\sqrt{\frac{3}{20}}\lambda_{35}
-\sqrt{\frac{3}{28}}\lambda_{48}+\frac{1}{\sqrt{7}}\lambda_{63}\right),\nonumber \\
\tilde{\bar{T}}^{34}&=&i\left(\frac{1}{\sqrt{10}}\lambda_{24}-\sqrt{\frac{3}{20}}\lambda_{35}
+\sqrt{\frac{3}{28}}\lambda_{48}-\frac{1}{\sqrt{7}}\lambda_{63}\right),\nonumber \\
\tilde{\bar{T}}^{56}&=&i\left(\frac{1}{\sqrt{10}}\lambda_{24}+\frac{1}{\sqrt{15}}\lambda_{35}
-\frac{2}{\sqrt{21}}\lambda_{48}-\frac{1}{\sqrt{7}}\lambda_{63}\right),\nonumber \\
\tilde{\bar{T}}^{13}&=&\frac{i}{2}\left( \lambda_{34}+\lambda_{62}
\right),\qquad \tilde{\bar{T}}^{23}=-\frac{i}{2}\left(
\lambda_{33}-\lambda_{61} \right),\qquad
\tilde{\bar{T}}^{14}=-\frac{i}{2}\left( \lambda_{33}+\lambda_{61} \right),\nonumber \\
\tilde{\bar{T}}^{24}&=&\frac{i}{2}\left( \lambda_{62}-\lambda_{34}
\right),\qquad \tilde{\bar{T}}^{45}=\frac{i}{2}\left(
\lambda_{58}-\lambda_{47} \right),\qquad
\tilde{\bar{T}}^{15}=\frac{i}{2}\left( \lambda_{59}-\lambda_{44} \right),\nonumber \\
\tilde{\bar{T}}^{25}&=&-\frac{i}{2}\left( \lambda_{45}+\lambda_{60}
\right),\qquad \tilde{\bar{T}}^{35}=\frac{i}{2}\left(
\lambda_{46}+\lambda_{57} \right),\qquad
\tilde{\bar{T}}^{16}=\frac{i}{2}\left( \lambda_{60}-\lambda_{45} \right),\nonumber \\
\tilde{\bar{T}}^{26}&=&\frac{i}{2}\left( \lambda_{44}+\lambda_{59}
\right),\qquad \tilde{\bar{T}}^{36}=\frac{i}{2}\left(
\lambda_{47}+\lambda_{58} \right),\qquad
\tilde{\bar{T}}^{46}=\frac{i}{2}\left( \lambda_{46}-\lambda_{57}
\right)
\end{eqnarray}
and
\begin{eqnarray}
\tilde{\bar{Y}}^{12}&=&-\frac{1}{2}\left(\lambda_{27}-\lambda_{16}+\lambda_{40}-\lambda_{55}\right),\qquad
\tilde{\bar{Y}}^{34}=-\frac{1}{2}\left(\lambda_{55}-\lambda_{16}+\lambda_{27}-\lambda_{40}\right),\nonumber \\
\tilde{\bar{Y}}^{56}&=&-\frac{1}{2}\left(\lambda_{55}-\lambda_{16}-\lambda_{27}+\lambda_{40}\right),\qquad
\tilde{\bar{Y}}^{13}=-\frac{1}{2}\left(\lambda_{54}-\lambda_{19}+\lambda_{26}-\lambda_{43}\right),\nonumber \\
\tilde{\bar{Y}}^{23}&=&-\frac{1}{2}\left(\lambda_{53}-\lambda_{18}-\lambda_{25}+\lambda_{42}\right),\qquad
\tilde{\bar{Y}}^{14}=\frac{1}{2}\left(\lambda_{18}+\lambda_{25}+\lambda_{42}+\lambda_{53}\right),\nonumber \\
\tilde{\bar{Y}}^{24}&=&-\frac{1}{2}\left(\lambda_{19}-\lambda_{26}-\lambda_{43}+\lambda_{54}\right),\qquad
\tilde{\bar{Y}}^{45}=-\frac{1}{2}\left(\lambda_{50}-\lambda_{23}+\lambda_{30}-\lambda_{39}\right),\nonumber \\
\tilde{\bar{Y}}^{15}&=&-\frac{1}{2}\left(\lambda_{31}-\lambda_{20}-\lambda_{36}+\lambda_{51}\right),\qquad
\tilde{\bar{Y}}^{25}=-\frac{1}{2}\left(\lambda_{21}+\lambda_{32}-\lambda_{37}-\lambda_{52}\right),\nonumber \\
\tilde{\bar{Y}}^{35}&=&-\frac{1}{2}\left(\lambda_{22}+\lambda_{29}+\lambda_{38}+\lambda_{49}\right),\qquad
\tilde{\bar{Y}}^{16}=-\frac{1}{2}\left(\lambda_{21}-\lambda_{32}-\lambda_{37}+\lambda_{52}\right),\nonumber \\
\tilde{\bar{Y}}^{26}&=&-\frac{1}{2}\left(\lambda_{20}+\lambda_{31}+\lambda_{36}+\lambda_{51}\right),\qquad
\tilde{\bar{Y}}^{36}=-\frac{1}{2}\left(\lambda_{50}-\lambda_{23}-\lambda_{30}+\lambda_{39}\right),\nonumber \\
\tilde{\bar{Y}}^{46}&=&-\frac{1}{2}\left(\lambda_{29}-\lambda_{22}+\lambda_{38}-\lambda_{49}\right).
\end{eqnarray}

\section{Scalar potential for $SO(4)\times USp(2)$ gauging}\label{potential_SO4_k2}
The scalar potential for compact gauge group $SO(4) \times USp(2)$
is given by
\begin{eqnarray}\label{eq:VSO4k2}
V&=&2 g_2^2 (3+\cosh b ) \sinh^6\frac{b}{2}+\frac{1}{16} g_1 g_2 \left[68+4 \cos(2 a_4)+2 \cos(2 (a_4-a_5))\right.\nonumber\\
&&\left.+4 \cos(2 a_5)+2 \cos(2 (a_4+a_5))+2 \cos(2 (a_4-a_6))+\cos(2 (a_4-a_5-a_6))\right.\nonumber\\
&&\left.+2 \cos(2 (a_5-a_6))+\cos(2 (a_4+a_5-a_6))+4 \cos(2 a_6)+2 \cos(2 (a_4+a_6))\right.\nonumber\\
&&\left.+\cos(2 (a_4-a_5+a_6))+2 \cos(2 (a_5+a_6))+\cos(2 (a_4+a_5+a_6))\right.\nonumber\\
&&\left.\left. +32 \cos^2 a_4  \cos^2 a_5  \cos^2 a_6  \cos(2 a_7)\right] (3+\cosh b ) \sinh^6\frac{b}{2}\right.\nonumber\\
&&\left.-4 g_1^2 \left[\cos^2 a_5  \cos^2 a_6  \cos^2 a_7  \cosh^2\frac{b}{2} (3+\cosh b )^2 \sin^2(2 a_4)\right.\right.\nonumber\\
&&\left.+64 \cos^2 a_4 \cosh^4\frac{b}{2} \sin^2 a_4  \sin^2 a_5 +64 \cos^2 a_4  \cos^2 a_5  \cosh^4\frac{b}{2}\right.\nonumber\\
 &&\left.\sin^2 a_4  \sin^2 a_6 +64 \cos^2 a_4  \cos^2 a_5  \cos^2 a_6  \cosh^4\frac{b}{2} \sin^2 a_4  \sin^2 a_7 \right.\nonumber\\
&&\left.+\frac{1}{16384}\left[51+259 \cos(2 a_4)+4 (-17+63 \cos(2 a_4)) \cosh b +(17+\cos(2 a_4)) \times\phantom{\frac{1}{2}}\right.\right.\nonumber\\
 &&\left.\cosh(2 b)+16 \cos^2 a_4  \cos (2 a_5) \sinh^4\frac{b}{2}+32 \cos^2 a_4  \cos^2 a_5  \cos(2 a_6) \sinh^4\frac{b}{2}\right.\nonumber\\
 &&\left.+64 \cos^2 a_4  \cos^2 a_5  \cos^2 a_6  \cos(2 a_7) \sinh^4\frac{b}{2}\right]^2+\frac{1}{2} \left[-4 \cos^4 a_4  \cos^2 a_5  \cos^2 a_6 \phantom{\frac{1}{2}}\right.\nonumber\\
&&\left.\cos^2 a_7  \sin^2 a_5 \sinh^6\frac{b}{2}-4 \cos^4 a_4  \cos^4 a_5  \cos^2 a_6  \cos^2 a_7  \sin^2 a_6  \sinh^6\frac{b}{2}\right.\nonumber\\
&&\left.-4 \cos^4 a_4  \cos^4 a_5  \cos^4 a_6  \cos^2 a_7  \sin^2 a_7  \sinh^6\frac{b}{2}-4 \sin^2(2 a_4) \sin^2 a_5  \sinh^2 b \right.\nonumber\\
&&\left.-16 \cos^2 a_4  \cos^2 a_5  \sin^2 a_4  \sin^2 a_6  \sinh^2 b -16 \cos^2 a_4  \cos^2 a_5  \cos^2 a_6  \sin^2 a_4  \right.\nonumber\\
&&\left.\sin^2 a_7  \sinh^2 b -\frac{1}{16} \cos^2 a_5  \cos^2 a_6  \cos^2 a_7  \sin^2(2 a_4) \left[7 \sinh\frac{b}{2}+3 \sinh\frac{3 b}{2}\right]^2\right.\nonumber\\
&&\left.-\frac{1}{4096}\left[16 \cos^2 a_4  \left[\cos(2 a_5)+2 \cos^2 a_5  \left(\cos(2 a_6)+2 \cos^2 a_6  \cos(2 a_7)\right)\right]\times \phantom{\frac{1}{2}}\right.\right.\nonumber\\
&&\cosh\frac{b}{2} \sinh^3\frac{b}{2}+2 [63 \cos(2 a_4)+17 \cosh b
-17] \sinh b
\nonumber\\
 & &\left.\left.\left.\phantom{\frac{1}{1}}+\cos(2
a_4) \sinh(2 b)\right]^2\right]\right].\label{V_SO4_k2}
 \end{eqnarray}


\end{document}